\documentclass[
    ,final            
  ]
  {aipproc}

\layoutstyle{8x11single}

\def\phibar{\left< \phi \right>}
\def\sigmav{\sigma_v}
\def\kms{km~sec$^{-1}$}

\def\hkpc{$h^{-1}$~kpc}

\def\Msun{M_\odot}
\def\Lstar{L^\ast}

\def\go{\mathrel{\raise0.35ex\hbox{$\scriptstyle >$}\kern-0.6em
\lower0.40ex\hbox{{$\scriptstyle \sim$}}}}
\def\lo{\mathrel{\raise0.35ex\hbox{$\scriptstyle <$}\kern-0.6em
\lower0.40ex\hbox{{$\scriptstyle \sim$}}}}

\begin{document}

\title{Constraints on Field Galaxy Halos from 
Weak Lensing and Satellite Dynamics}

\author{Tereasa G.\ Brainerd}{
  address={Boston University, Institute for Astrophysical Research,
725 Commonwealth Ave., Boston, MA 02215}
}

\begin{abstract}
Here I summarize constraints on the nature of the dark
matter halos of field galaxies that have been obtained from
the most
recent investigations of (i) weak galaxy--galaxy lensing and (ii) the
dynamics of satellite galaxies in orbit about large host
galaxies.  Both of these techniques
are statistical in their nature (i.e., large samples of
galaxies are required to obtain a ``signal''), but since they have
inherently different selection biases and systematic errors,
they are quite complementary to each other.  Results of work over
the last several years on weak lensing and satellite dynamics
is revealing a remarkably consistent picture regarding
the dark matter halos of bright field galaxies ($L \go \Lstar$).  The halos
extend to large physical
radii ($\go 150~h^{-1}$~kpc) and are flattened in
projection on the sky, there is a marked difference in
the depths of the potential wells of early--type galaxies and 
late--type galaxies, and the velocity dispersion profiles of
the halos, $\sigma_v(r_p)$, decrease at large projected radii.
All of these are expected to hold true in
a cold dark matter universe and, while neither
technique can address the possible small--scale ($\lo 5~h^{-1}$~kpc)
conflicts between cold dark matter and observed galaxies, on
scales $\go 50~h^{-1}$~kpc both techniques yield results that are 
consistent with each other, and with the predictions of cold dark matter.
\end{abstract}

\maketitle


\section{Introduction}

The existence of dark matter halos surrounding large, bright
galaxies is well established (e.g., \cite{Mathews:Brighenti},
\cite{Sofue:Rubin}, \cite{Fich:Tremaine}, \cite{deZeeuw:Franx} and
references therein), and in the standard cold dark matter (CDM)
paradigm, the halos of large field galaxies are expected to extend
to virial radii of $\sim 100h^{-1}$~kpc to $\sim 200h^{-1}$~kpc
and have masses of $\sim 10^{12} h^{-1} \Msun$ (e.g., \cite{NFW:1995},
\cite{NFW:1996}, \cite{NFW:1997}).  Until very recently, however, direct 
observational constraints on the nature of the dark matter halos
of field galaxies have not been especially strong.  In particular,
it has been challenging to address
the question as to whether the halos of observed galaxies are 
consistent with the halos that one would expect in a CDM universe.

The lack of a Keplerian fall--off in the rotation curves of the
disks of most spiral galaxies (e.g., \cite{Sofue:Rubin}) indicates
that the dark matter halos extend far beyond the visible radii
of the galaxies.  Therefore, in order to place constraints on the
total mass distribution, it is necessary to use tracers of the halo
potential that exist at large projected radii
($\go 100h^{-1}$~kpc).  Two such tracers
of the large--scale potential
are satellite galaxies that are in orbit about isolated host galaxies, and
photons 
emitted by distant galaxies that, on their way to the observer,
happen to pass through the
potential wells of more nearby galaxies at small impact parameters
(i.e., gravitational lensing).  ``Strong'' gravitational lensing, in
which multiple and highly--distorted images of a source occur, is
a rare phenomenon because it requires nearly perfect alignment of
the lens and source galaxy (e.g., \cite{dasBuch}).  ``Weak'' gravitational
lensing, in which multiple images and significant image
distortion do not occur, is, however, commonplace in the universe
(e.g., \cite{Yannick},
\cite{Matthias:Peter}, \cite{Wittman}, \cite{Matthias:Ramesh}),
and it is on this extremely mild regime of gravitational lensing
that I will focus for this discussion.  

Weak lensing of background galaxies by foreground galaxies 
(``galaxy--galaxy'' lensing) and the motions of satellite galaxies
about host galaxies are phenomena that can only lead to constraints
on halo potentials through ensemble averages over statistically
large samples.  That is, for any given foreground galaxy, the 
distortion that it induces in the images of background galaxies 
due to weak lensing is so small that the signal cannot be
detected convincingly for any one foreground lens galaxy.  Similarly,
isolated host galaxies are typically found to have 1 to 2
satellite galaxies on average and, so, the potential of any one host galaxy
cannot be constrained at all well by the motions of its own
satellites.  Both galaxy--galaxy lensing and satellite dynamics,
therefore, lead to statistical constraints on the halo population
as a whole and by their very nature they require large samples of 
galaxies in order to obtain such constraints. 
Until several  years ago, galaxy--galaxy lensing and satellite
dynamics were both tantalizingly close (or frustratingly close,
depending on one's point of view) to being able to fulfill their
theoretical promise to map out the gravitational potentials of
the halos of field galaxies.  With the advent of routine availability
of wide--field imagers and the completion (or
near completion) of large redshift surveys, however, both galaxy--galaxy
lensing and satellite dynamics are now yielding sufficiently strong
constraints on the dark matter halos of galaxies that the observations
can be used to test the theoretical predictions
(i.e., CDM) at a substantive level.

\begin{figure}
\centerline{
\scalebox{0.9}{%
\includegraphics{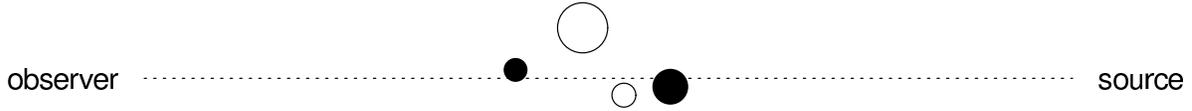}%
}
}
\caption{
Schematic representation of multiple lenses along the line of sight
to a given source galaxy.
}
\label{multiple}
\end{figure}

There are distinct advantages and disadvantages of galaxy--galaxy
lensing versus satellite dynamics when it comes to constraining halo
potentials.  A clear advantage of galaxy--galaxy lensing is that
it can be applied to {\it all} foreground galaxies and, since
gravitational lensing is affected only by the total  mass along the
line of sight and not its dynamical state, the halos of the
foreground lens galaxies need not be virialized.  
A complicating factor in galaxy--galaxy lensing is that it is not
correct to assume that each background galaxy has been lensed solely
by one foreground galaxy (e.g., \cite{BBS}, \cite{Brainerd:Blandford},
\cite{TGB:Lausanne}).  Instead, photons emitted by the distant galaxies
are deflected by {\it all} mass along the line of sight, including 
individual galaxies, groups, and clusters (e.g., Figure~\ref{multiple}).
That is, galaxy--galaxy lensing is inherently a multiple--deflection
problem and care must be taken when using observations of galaxy--galaxy
lensing to constrain the halos of a given subset of lens galaxies
(i.e., the halos of early--type galaxies versus late--type galaxies, or
the halos of high--luminosity galaxies versus low--luminosity galaxies).
Therefore, a computation of the weak lensing signal about the white
lenses in Figure~\ref{multiple} above is {\it not} identical to a measurement
of weak lensing signal {\it produced by} the white lenses since the
black lenses also contribute to the net shape of the final image.  That
is not to say that galaxy--galaxy lensing cannot be used to probe
the potentials of halos surrounding lenses of differing types; it
most certainly can, but the presence of multiple deflections in the
data must be taken into account when modeling the observed signal.  
In the case of relatively shallow
data ($z_{\rm lens} \sim 0.15$), most sources will have been lensed
by only one foreground galaxy (e.g., \cite{Guzik:Seljak}), but in
deep data sets ($z_{\rm lens} \sim 0.5$) most source galaxies will
have been lensed at a significant and comparable level by two or 
more foreground
galaxies (e.g., \cite{BBS}, \cite{TGB:Lausanne}).
A further disadvantage of galaxy--galaxy lensing is that the 
signal is very small (systematic image distortions of $\lo 1$\% or
in the image ellipticities), so the images of millions of background
galaxies must be obtained and, in general, be meticulously
corrected for the 
presence of anisotropic, spatially--varying point spread functions.
Finally, it is possible that Newtonian tidal distortions of genuine
satellites of the lens galaxies could masquerade as a weak lensing
signal.  Happily, such distortions appear to be at most a very small
contributor to the observed weak lensing signal (e.g., 
\cite{BBS}, \cite{Tony:Nature}.
\cite{Bernstein:Norberg}, \cite{Hirata}). 

An advantage to using dynamics of satellite galaxies to probe the
potentials of the halos of isolated host galaxies is that, unlike
deep weak lensing data, the only important potential well in the
problem is that of the host galaxy.  In principle, this is a ``cleaner'',
more straightforward
probe of the halo potential which is intentionally restricted to the
physical scales that one would expect to characterize the halos of
individual large galaxies.  However, there are a number of
arguments against the use of 
satellite dynamics to probe the mass distributions of host galaxies:
(1) satellites must be found at large projected radii in order to
probe the halo potential on the very largest scales, (2) noise  is 
introduced by the presence of ``interlopers''
(i.e., galaxies that are selected as satellites but which are, in fact,
not associated dynamically
with the host galaxy), and (3) the relaxation times
of these systems are large compared to the age of the universe.  The
first argument is  much less compelling now than it
was in the past simply because of 
the availability of large redshift surveys (i.e., the
data bases are now sufficiently big that although it is rare to
find satellites at, say, a projected radius of 500~\hkpc,  the large number
of redshifts that are now available makes it possible to compile statistically
significant samples).  The second argument has also become much
less compelling 
with the realization that it is straightforward to
account for the effects of interloper galaxies on the determination of the
velocity dispersion (see below).  
The third argument can still be compelling, since 
it makes little sense to apply a
virial--type  mass estimator to systems which are not relaxed.  However,
an assumption of virialization is not necessary {\it a priori}, and the
use of secondary infall models can be used to bypass this assumption
(e.g., \cite{ZSFW}).

For the sake of a certain amount of brevity and, at the very least,
an attempt at providing some level of coherent argument, I will
focus here on only the most recent results that are directly
relevant to the halos of
field galaxies and which have been obtained from four
large surveys:  the COMBO--17 Survey, the Red--Sequence Cluster Survey (RCS),
the Sloan Digital Sky Survey (SDSS), and
the Two Degree Field Galaxy Redshift Survey (2dFGRS).
Even with
this restriction, it is simply not possible to discuss all of the most recent
results from these studies in great detail, and the reader should consult the
source literature for further information.
Finally, I should hasten to add that all errors or omissions in
this article in regards to my colleagues' work are entirely unintentional
and entirely my own fault.  I can only hope that my colleagues will be
kind enough to forgive me.

\section{The Surveys}

\subsection{COMBO--17: Galaxy--Galaxy Lensing}

The acronym COMBO--17 stands for ``Classifying Objects by Medium--Band
Observations in 17 filters'' \cite{Wolf:2001}, \cite{Wolf:2003}, 
\cite{Wolf:2004}.  The COMBO--17 survey consists of high--quality
imaging data with the ability to obtain both rest frame colors and
accurate photometric redshifts ($\delta z/(1+z) < 0.01$ for
$R < 21$, 
$\delta z/(1+z) \sim 0.02$ for $R \sim 22$, and
$\delta z_{\rm phot} < 0.1$ for $R < 24$).  The survey consists of
5 fields, including an extended region in the location
of the Chandra Deep Field South (CDFS).  The observations were carried
out using the Wide Field Imager
at the 2.2--m MPG/ESO telescope.  The field of view of the camera
is $34' \times 33'$ and the 17--band filter set covers a wavelength
range of $350~{\rm nm} \lo \lambda_{\rm obs} \lo 930~{\rm nm}$.  The
latter allows for a rough determination of the spectral energy distributions
of the objects, which in turn leads to both reliable classification of 
the objects into galaxies, quasars, and stars, as well as the ability
to determine accurate photometric redshifts. A catalog containing
astrometry, photometry in all 17 bands, object classification, and
photometric redshifts for the
63,501 objects in the extended CDFS is publicly--available \cite{Wolf:2004}
(see \url{http://cdsweb.u-strasbg.fr/cgi-bin/qcat?J/A+A/421/913}).
The COMBO--17 results that will be discussed here 
consist of efforts to use galaxy--galaxy 
lensing to study dark matter halos.
The data set is particularly 
well--suited to this task because of the reliability with which
background galaxies (i.e., lensed sources) can be separated from
foreground galaxies (i.e., the lenses).  Note, too, that
although the full COMBO--17 survey covers 5
fields, the results shown here come from only 3 of the fields
(a field centered on the cluster A901, the CDFS field, and
a random field \cite{combo17:old}).

\subsection{RCS: Galaxy--Galaxy Lensing}

The RCS (\cite{RCS1}, \cite{RCS2}) is a somewhat shallow
(5$\sigma$ point source detection limits of $R_C \sim 24.8$
and $z' \sim 23.6$), wide field ($\sim 90$~sq.\ deg.) imaging
survey that was designed primarily to search for galaxy clusters
out to redshifts of $z \sim 1.4$.  The images for the complete
survey
were obtained with the CFHT and CTIO 4--m telescopes using
mosaic cameras, and consist of 22 widely--separated patches of
$\sim 2.1^\circ \times 2.3^\circ$.   The RCS results that will
be discussed here consist of galaxy--galaxy lensing
studies and were obtained
from $\sim 42$~sq.~deg.\ of northern RCS data.  
Without spectroscopic
or photometric redshift information, the RCS galaxy--galaxy results
had to be obtained from a rough separation of lenses and sources 
that was based
upon apparent magnitude cuts (i.e., galaxies with ``faint'' 
apparent magnitudes are on
average background objects while galaxies with ``bright'' apparent
magnitudes are on average foreground objects).  Although the 
foreground--background distinction between a given pair of galaxies in
the RCS data is by no means as secure as in the COMBO--17 data, the
RCS is nevertheless a superb data set for galaxy--galaxy studies simply
because of the area covered ($\sim 45$ times larger than COMBO--17
for the weak lensing work).
Given that weak lensing is primarily a statistical game, this is 
a good example of how well the galaxy--galaxy lensing signal can
be detected and also used to constrain the nature of dark
matter halos given only minimal distance information and
a tremendous number of candidate lenses and sources.

\subsection{SDSS: Galaxy--Galaxy Lensing \& Satellite Dynamics}

The SDSS is a combined photometric and spectroscopic survey that
will ultimately map roughly one quarter of the sky
above $l \sim 30^\circ$ and provide redshifts of
$\sim 10^6$ galaxies and $\sim 10^5$ quasars
with $r' \lo 17.8$.  The SDSS is a 
fully--digital survey and makes use of
5 broad optical bands ($u'$, $g'$, $r'$, $i'$, $z'$) for 
photometry.  The data for the SDSS are being acquired at the
Apache Point Observatory in Sunspot, New Mexico using a 2.5--m
telescope, as well as three, smaller subsidiary telescopes for the
purposes of photometric calibration, monitoring of the seeing,
and scanning for clouds.  The rms galaxy redshift errors
are $\sim 20$~\kms to $\sim 30$~\kms (e.g., \cite{Prada}, \cite{Max}).
A technical summary
of the SDSS can be found in York et al.\ \cite{York:2000}, information about
the main galaxy sample is given by Strauss et al.\ \cite{Strauss:2002},
and information
about the photometric system and photometric calibration is
given by Fukugita et al.\ \cite{Fukugita:1996},
Hogg et al.\ \cite{Hogg:2001}, and Smith et al.\ \cite{Smith:2002}.
All of the SDSS data, including astrometry, photometry, redshifts,
and spectra, are available via the SDSS website
(\url{http://www.sdss.org}) using
structured queries that can search and combine the individual data bases.
The third SDSS data release occurred on September~27, 2004 and includes
spectra of 374,767 galaxies, spectra of 51,027 quasars, and photometry
of 141 million unique objects.
The SDSS results that will be discussed here 
consist of both galaxy--galaxy lensing studies
and studies of the satellites of large, isolated
galaxies.

\subsection{2dFGRS: Satellite Dynamics}

The 2dFGRS is a spectroscopic survey in which the target objects
were selected in the $b_J$ band from the  Automated Plate Measuring
(APM) galaxy survey (\cite{Maddox:1990a},
\cite{Maddox:1990b}) and extensions  to
the original survey.  
A detailed discussion of the survey and the data base is given by
Colless et al.\ \cite{Colless:2001}.
The observations, which are now complete,
were carried out at the Anglo--Australian Telescope
using the Two Degree Field (2dF) multifiber spectrograph.
The  final data release occurred on June 30, 2003
\cite{Colless:2003}
and  includes reliable redshifts of 221,414  galaxies
with extinction corrected magnitudes of
$b_J \ge 19.45$, covering  an area over $\sim 1500$  square  degrees. 
Galaxies with reliable redshifts have an rms uncertainty
of 85~\kms \cite{Colless:2001}.
All data, including spectroscopic catalogs (245,591 objects),
photometric catalogs (382,323 objects), and FITS files containing
the spectra, are
publicly--available from the  2dFGRS website
(\url{http://msowww.anu.edu/au/2dFGRS}).
The 2dFGRS data base is fully--searchable via structured queries, and
on--line documentation is available on the 2dFGRS website.
The  photometric transformation from the
SDSS band passes to $b_J$ is
\begin{equation}
b_J = g' + 0.155 + 0.152(g' -  r')
\label{transform}
\end{equation}
(e.g., \cite{Norberg:2002}).
The 2dFGRS results that will be discussed here 
consist of
investigations into the nature of the dark matter halos of large,
isolated galaxies  that are orbited by one or more satellite
galaxies.

\section{Probing Halo Potentials with Weak Lensing}

General Relativity tells us that any mass will cause a curvature
of spacetime in its vicinity.  Therefore, any mass located
along the line of sight to a distant luminous object will act as
a gravitational lens by deflecting light rays emanating from the object
as they propagate through the universe.  The
most striking instances of gravitational lensing (e.g.,
multiple images, rings, arcs) are
examples of rare phenomena caused by
strong gravitational lenses, which greatly distort the images of distant
galaxies.  In contrast to this,
weak gravitational lenses distort the images of distant galaxies
very little but produce a net coherent pattern of image distortions in
which there is a slight preference for the lensed galaxies to be
oriented tangentially with respect to the direction vector that
connects their centroids with the center of the gravitational
potential of the lens.
While weak lenses do not give rise to stunning individual
images, they are detectable in a statistical sense via ensemble
averages over many mildly--distorted images
(e.g., \cite{Yannick},
\cite{Matthias:Peter}, \cite{Wittman}, \cite{Matthias:Ramesh}).

Provided the distance traveled by the light ray is very much
greater than the scale size of the lens, it is valid to
adopt the ``thin lens approximation'' in order to describe
a gravitational lens.  Consider a lens with an arbitrary
3-dimensional potential, $\Phi$.  In the thin lens approximation
a conveniently scaled 2-dimensional potential for
the lens (i.e., the 3-dimensional potential of the lens integrated
along the optic axis) is given by
\begin{equation}
\psi(\vec{\theta}) = \frac{D_{ls}}{D_l D_s} \; \frac{2}{c^2}
\int \Phi(D_d \vec{\theta},z) dz,
\end{equation}
where $\vec{\theta}$ is the location of the lensed image on the
sky, measured with respect to the optic axis, and $D_{ls}$, $D_l$, and
 $D_s$ are angular diameter distances 
between the lens and source, observer and lens, and observer and
source, respectively
(e.g., \cite{dasBuch}).
It is then straightforward to relate the gravitational potential of the
lens 
to the two fundamental quantities that
characterize the lens: the convergence ($\kappa$) and
the shear ($\vec{\gamma}$).  The convergence, which
describes the isotropic focusing of light rays, is given by
\begin{equation}
\kappa(\theta) = \frac{1}{2} \left( \frac{\partial^2 \psi}
{\partial \theta_1^2} + \frac{\partial^2 \psi}{\partial \theta_2^2}
\right).
\end{equation}
The shear describes tidal gravitational forces acting across
a bundle of light rays and, therefore, the shear
has both a magnitude, $\gamma = \sqrt{\gamma_1^2 + \gamma_2^2}$,
and an orientation, $\varphi$.  In
terms of $\psi$, the components of the
shear are given by
\begin{equation}
\gamma_1 (\vec{\theta}) = \frac{1}{2} \left( \frac{\partial^2 \psi}
{\partial \theta_1^2} - \frac{\partial^2 \psi}{\partial \theta_2^2}
\right) \equiv \gamma(\vec{\theta}) \cos \left[ 2\varphi(\theta) \right]
\end{equation}
and
\begin{equation}
\gamma_2 (\vec{\theta}) = \frac{\partial^2 \psi}{\partial \theta_1
\partial \theta_2} = \frac{\partial^2 \psi}{\partial \theta_2 
\partial \theta_1} \equiv \gamma(\vec{\theta}) \sin \left[
2 \varphi(\vec{\theta}) \right].
\end{equation}

The effect of convergence and shear acting together in a gravitational
lens is to distort the images of distant objects.  Consider
a source galaxy which is spherical in shape.  In the absence of a
gravitational lens, an observer would see an image of the galaxy which
is truly circular.  If a gravitational lens is interposed along the
line of sight to the distant galaxy,
the observer will see an image which, to first order,
is elliptical and the major axis
of the ellipse will be oriented tangentially with respect to the
direction vector on the sky that connects the centroids of the image
and the lens.  That is, the circular source is distorted into an
ellipse, and to first order the distortion consists of both a
tangential stretch of $(1 - \kappa - \gamma)^{-1}$ and a radial
compression of $(1 - \kappa + \gamma)^{-1}$ (e.g., \cite{dasBuch}).
In the weak lensing regime, both the convergence and shear are
small ($\kappa << 1$
and $\gamma  <<1$).

The fundamental premise in all attempts to detect weak lensing is
that, in the absence of lensing, galaxy images have an intrinsically
random ellipticity distribution.  Gravitational lensing then introduces
a shift in the ellipticity distribution that, in the mean,
manifests as a tangential
alignment of background sources around foreground lenses.  The image
of a distant galaxy can be approximated an ellipse with complex 
image ellipticity given by
\begin{equation}
\epsilon =  \frac{a^2 - b^2}{a^2 + b^2} e^{2i\phi} = \epsilon_1 + i \epsilon_2,
\end{equation}
where $a$ and $b$ are the major and minor axes, respectively, and $\phi$ is
the position angle.  The complex image ellipticity is often referred to
as the ``image polarization'' (e.g., \cite{BSBV}) and is computed in
terms of flux--weighted second moments,
\begin{equation}
Q_{i,j} = \sum_{i,j} I_{i,j} W_{i,j} x_i x_j,
\end{equation}
where $I_{i,j}$ is the intensity at a given pixel and $W_{i,j}$ is a
weighting function.  The real and imaginary components of the image polarization
are then given by:
\begin{equation}
\epsilon_1 = \frac{Q_{1,1} - Q_{2,2}}{Q_{1,1} + Q_{2,2}}, \;\;\;\;\;\;
\epsilon_2 = \frac{2Q_{1,2}}{Q_{1,1} + Q_{2,2}}.
\end{equation}

The {\it observed} image polarization for any one source is, of course, a
combination of its intrinsic ellipticity and any ellipticity that is induced
by lensing.  In the limit of weak lensing, the observed image polarization,
$\epsilon^{\rm obs}$, is
related to the intrinsic image polarization,
$\epsilon^{\rm int}$ through a shift in the complex
plane.  Although we cannot determine $\epsilon^{\rm int}$ for any one
particular source galaxy,
we have that the mean intrinsic ellipticity distribution for an 
ensemble of source galaxies is $\left< \epsilon^{\rm int} \right> = 0$ since
the galaxies should be randomly--oriented in the absence of lensing.  An 
estimator for the shear induced by weak lensing is then
$\gamma = \left< \epsilon^{\rm obs} \right> / 2$ (e.g., \cite{BSBV}).  
This simple estimator does
not reflect the fact that the way in which the shear alters the shape of a source
depends upon its intrinsic ellipticity, and in practice this is generally
taken into account when computing the shear.  
See, e.g., \cite{KSB}, 
\cite{Bernstein:Jarvis}, and \cite{Sheldon} for discussions of the 
``shear polarizability'' and ``shear responsivity'' of sources.  In addition, it
is worth noting that, while it is common practice to approximate image
shapes as ellipses, there will be some images that  have been sufficiently
distorted by galaxy--galaxy lensing that a mild bending, or ``flexion'',
of the images will occur and such images cannot be accurately 
represented as ellipses.  In principle, flexion of images can
be used to detect weak lensing with a signal--to--noise that is increased
over the common practice of fitting equivalent image ellipses
\cite{Dave}, \cite{Dave:Priya}.  A preliminary
application of this technique \cite{Dave} has been carried out with the Deep Lens
Survey \cite{DLS}, and it will be interesting to see how the technique is further
developed and implemented in practice.

The first attempts to detect systematic weak lensing of 
background galaxies by foreground galaxies (\cite{TVJM},
\cite{Tony}) were met with a certain degree of skepticism
because the apparent distortion of the source galaxy images
was rather smaller than one would expect based upon the typical
rotation velocities of the disks of large spiral galaxies.  The situation
changed when Brainerd, Blandford \& Smail \cite{BBS} measured
the orientations of 506 faint galaxies ($23 < r_f \le 24$)
with respect to the locations of 439 bright 
galaxies ($20 \le r_b \le
23$) and found that the orientation of the faint
galaxies was inconsistent with a random distribution at
the 99.9\% confidence level.  The faint galaxies
showed a clear preference for tangential alignment with the
direction vector on the sky that connected the centroids of
the faint and bright galaxies, in agreement with the expectations
of systematic weak lensing of the faint galaxies by the bright galaxies.  

\begin{figure}
\centerline{
\scalebox{0.60}{%
\includegraphics{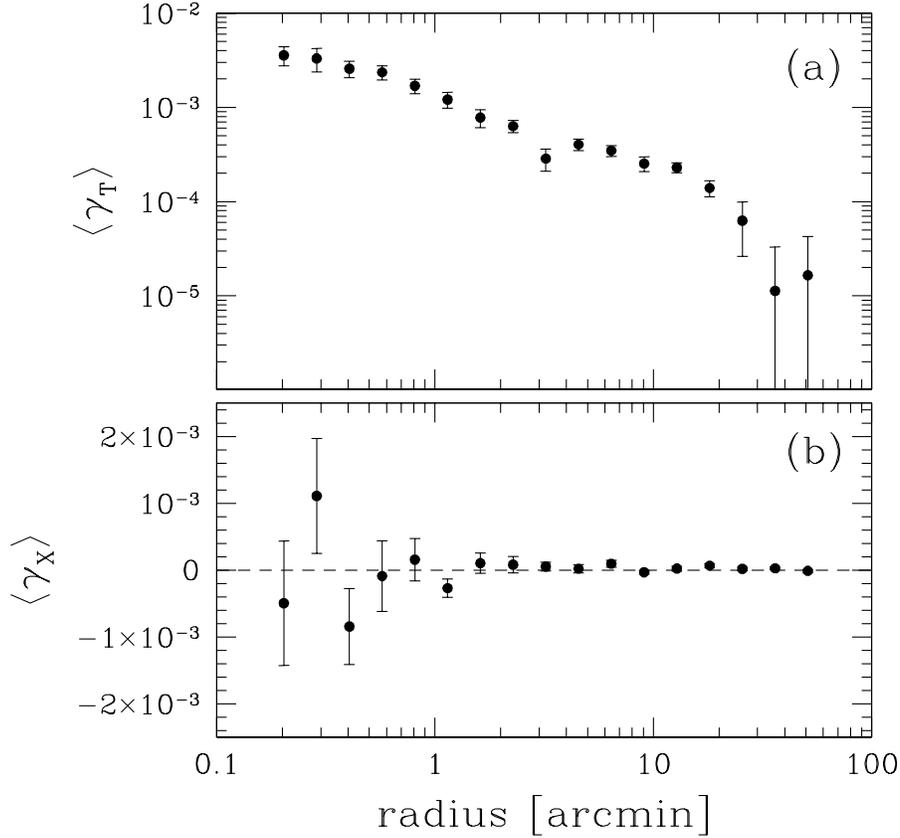}%
}
}
\caption{
a) Mean tangential shear computed about the lens centers in 
$\sim 42$~sq.~deg.\ of the RCS \cite{Henk:2004}.
Here foreground galaxies and background galaxies have been separated on
the basis of apparent magnitude alone.  Bright, lens galaxies have
$19.5 < R_C < 21$ and faint, source galaxies have $21.5 < R_C < 24$.
b) Same as in a) except that here each background galaxy image
has been rotated by
$45^\circ$.  This is a control statistic
and in the absence of systematic errors it should be consistent
with zero on all scales.
Figure kindly provided by Henk Hoekstra.
}
\label{Henk:gammat}
\end{figure}

Almost immediately, a number of similar investigations followed in
the wake of Brainerd, Blandford \& Smail \cite{BBS} (\cite{Griffiths},
\cite{dellAntonio}, \cite{Hudson}, \cite{Ebbels},  \cite{Fischer},
\cite{Hoekstra}, \cite{Jaunsen}).  These studies
made use of a wide variety of data and analysis techniques,
and all were broadly consistent with one another and with the
results of Brainerd, Blandford \& Smail \cite{BBS} (see, e.g., 
the review by Brainerd \& Blandford
\cite{Brainerd:Blandford}).  
The first truly undeniable detection of galaxy--galaxy lensing was obtained
by Fischer et al.\ \cite{Fischer} with 225 sq.\ deg.\ of early commissioning
data from the SDSS, and it was this result in particular that helped to
make the study
of galaxy--galaxy lensing into a respectable endeavor, whereas previously
many had considered the whole field rather dodgy at best.    
Fisher et al.\ \cite{Fischer} demonstrated conclusively that 
even in the limit of somewhat poor
imaging quality, including the presence of an anisotropic
point spread function due to drift scanning, galaxy--galaxy
lensing can be detected with very high significance in wide--field
imaging surveys.
In the last few years, detections of galaxy--galaxy lensing
and the use of the signal to constrain the dark matter halos
of field galaxies has improved dramatically (\cite{combo17:old},
\cite{Sheldon}, \cite{Henk:2004},
\cite{McKay:lens}, \cite{Deano}, \cite{Gillian},
\cite{Henk:2003}, 
\cite{combo17:new}, \cite{Uros}) owing to
a number of factors that include such things as
very large survey areas, sophisticated
methods for correcting image shapes due to anisotropic and
spatially--varying point spread functions, and the use
of distance information for large numbers foreground lens galaxies 
in the form of either spectroscopic or photometric redshifts.  

Figure~\ref{Henk:gammat} shows one
example of the high statistical significance with which weak lensing
due to galaxies is now being routinely detected.  The
result comes from an analysis of the distortion of the images of
$\sim 1.5\times 10^6$ source galaxies due to $\sim 1.2\times 10^5$ lens
galaxies
in the RCS \cite{Henk:2004}, where the lens and source populations were
separated solely on the basis of their apparent magnitudes.  
The top panel of Figure~\ref{Henk:gammat}
shows the mean tangential
shear {\it computed} about the lens centers which, because of the clustering
of the lens galaxies, is not simply interpreted as the tangential
shear {\it due to} individual lens centers.  Instead, it is a 
projected (i.e., 2--dimensional) galaxy--mass
cross--correlation function, and in order to compute the average
properties of the halos of the lens galaxies it is necessary to, e.g., 
make use of Monte Carlo simulations that include all of the multiple
weak deflections that the sources have undergone. The bottom panel of
Figure~\ref{Henk:gammat} shows a control statistic in which the tangential
shear about the lens centers is computed after rotating the images of
the sources by $45^\circ$.  If the signal in the top panel of 
Figure~\ref{Henk:gammat} is caused by gravitational lensing, the control
statistic in the bottom panel of Figure~\ref{Henk:gammat} should be consistent
with zero (and indeed it is).  Note that, although the tangential shear
about the RCS lenses persists to scales of order $0.5^\circ$, the shear
on such large scales is not indicative of the masses of individual lens
galaxies; rather
it reflects the intrinsic clustering of the lenses.
It is also worth noting that less than decade ago observers were struggling
to measure a tangential shear of $\lo 0.01$ with a modest degree of
confidence. Now, 
however, confident detection of tangential shears of $\lo 0.0001$ is
effectively ``routine'' in these extremely large data sets.

\begin{figure}
\centerline{
\scalebox{0.75}{%
\includegraphics{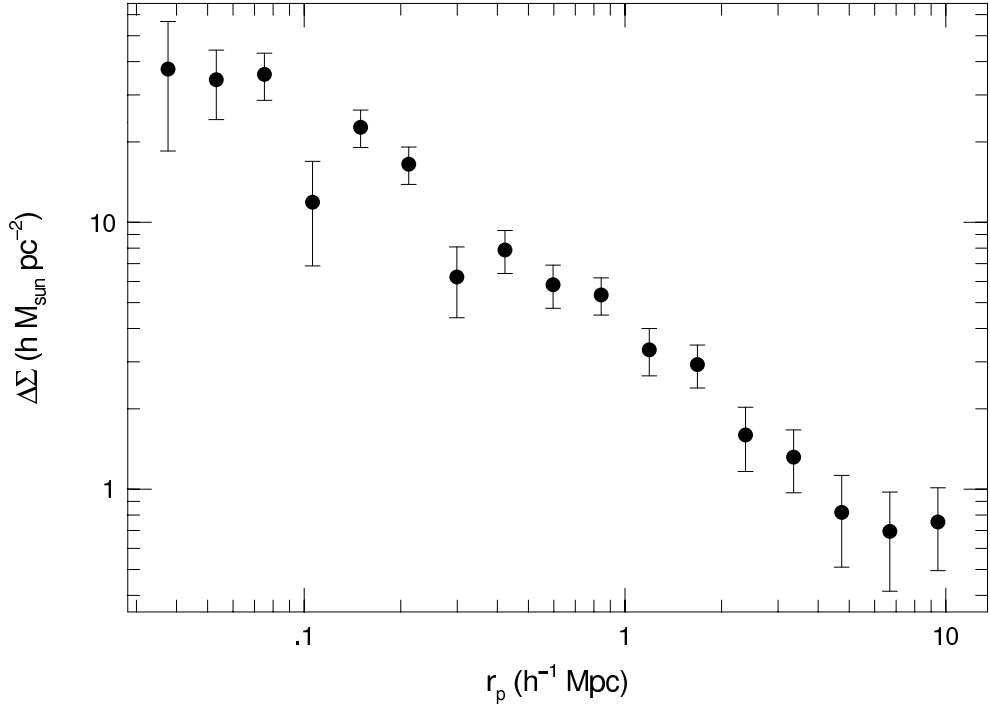}%
}
}
\caption{
Mean excess projected mass density around weak galaxy lenses in the
SDSS \cite{Sheldon}.  Here $\sim 1.27\times 10^5$ lenses with spectroscopic
redshifts and $\sim 9.0\times 10^9$ sources with photometric redshifts
have been used in the calculation.  The values of $\Delta \Sigma (r_p)$ shown
in this figure have been corrected for the clustering of the sources
around the lenses.  Data kindly provided by Erin Sheldon.
}
\label{Erin:fig6}
\end{figure}

The mean tangential shear, $\gamma_T(r_p)$, in an annulus of 
projected radius $r_p$
is related to the projected surface mass density of the lens through
\begin{equation}
\Sigma_c \gamma_T(r_p) = \overline{\Sigma}( r < r_p) - 
\overline{\Sigma}(r_p) \equiv \Delta \Sigma (r_p),
\label{Delta:Sigma}
\end{equation}
where $\overline{\Sigma}( r < r_p)$
is the mean surface mass density interior to the projected radius $r_p$,
$\overline{\Sigma}(r_p)$ is the projected surface mass density
at radius $r_p$ (e.g., \cite{Jordi:1991}, \cite{Kaiser:1994}, 
\cite{Gillian}), and $\Sigma_c$ 
is the so--called critical surface
mass density:
\begin{equation}
\Sigma_ c \equiv \frac{c^2 D_s}{4\pi G D_l D_{ls}},
\label{Sigmac}
\end{equation}
where $c$ is the velocity of light and
$D_s$, $D_l$, and $D_{ls}$ are again angular diameter distances \cite{dasBuch}.
The quantity $\Delta \Sigma (r_p)$ above
is, therefore, a mean excess projected mass
density.
Shown in Figure~\ref{Erin:fig6} is the mean excess projected surface
mass density in physical units of $h~M_\odot~{\rm pc}^{-2}$ for 
$\sim 1.27\times 10^5$ lens galaxies in the SDSS for which spectroscopic
redshifts are known \cite{Sheldon}.  In addition to spectroscopic 
redshifts for the lenses, photometric redshifts were used for 
$\sim 9.0\times 10^9$ source galaxies.  Moreover, because the redshifts
of the lens galaxies are known, $\Delta \Sigma(r_p)$ can be computed
as a function of the physical projected radius at the redshift of the
lens (rather than an angular scale).  In Figure~\ref{Erin:fig6}, 
$\Delta \Sigma(r_p)$ has been corrected for the clustering of the sources
around the lenses
via a function which is effectively a weighted cross--correlation 
function between the lenses and sources \cite{Sheldon}.

Having obtained a measurement of $\gamma_T (\theta)$, or equivalently 
$\Delta \Sigma (r_p)$, constraints can then be placed on the nature
of the dark matter halos of the lens galaxies by modeling the 
observed signal.  As mentioned earlier,
quite a bit of care has to be taken in doing this
if the goal is to constrain the halo parameters as
a function of, say, the host luminosity, color, or morphology (see, e.g.,
\cite{Guzik:Seljak}).  In the past few years, however,  
good constraints on the mass of an ``average'' halo associated
with an $L^\ast$ galaxy, as well as fundamental differences between
the halos of $L^\ast$ ellipticals versus $L^\ast$ spirals, have
emerged from galaxy--galaxy lensing studies and it is those studies
which are summarized below.

\section{Probing Halo Potentials with Satellite Galaxies}

In order to use satellite galaxies to probe the potentials of
host galaxies, one needs to {\it define} an appropriate sample
of host and satellite galaxies.  Unlike cosmology simulators who are
blessed with full 6-dimensional phase space information, observers
are, of course, limited to 3 dimensions (RA, DEC, and redshift).
Given this limited information, then, one must base the selection
criteria on projected radii (evaluated at the redshift of the host)
and relative radial velocities, $dv$, of the candidate
hosts and satellites.
To guarantee that the dynamics of the satellites are
determined solely by their host galaxy, the hosts must be 
determined to be ``isolated'' in some sense.  That is, if another
large, bright galaxy is too close to a candidate host galaxy
to guarantee that the satellite orbits are affected solely
by the candidate host, that candidate host and its satellites are rejected
from the sample.  Satellites must, necessarily, be fainter than
their host, be found within some reasonable projected radius
of the host, and have some reasonable line of sight velocity with
respect to the host.

There are a number of different selection criteria that have
been used in the recent literature, and three sets of
selection criteria that
have been used in more than one investigation are
summarized below:

\begin{enumerate}
\item
Hosts must be at least 8 times brighter than any other
galaxy that is within $r_p < 500$~kpc and
$|dv| < 1000$~km~sec$^{-1}$.
In addition, hosts must be at least 2 times brighter than any
other galaxy that is within $r_p < 1$~Mpc and
$|dv| < 1000$~km~sec$^{-1}$.
Satellites must be at least 8 times fainter than their host,
must be found within $r_p < 500$~kpc, and
must have $|dv| < 500$~km~sec$^{-1}$.  Here $h =0.7$ has
been adopted (\cite{ZSFW}, 
\cite{TGB:holmberg}).

\item
Hosts must be
at least 2 times brighter than any other galaxy that
falls within
$r_p < 2.86$~Mpc and
$|dv| <  1000$~km~s$^{-1}$.
Satellites must be at least 4 times fainter
than their host, must be found within
$r_p < 714$~kpc
and
must have $|dv| < 1000$~km~s$^{-1}$.  Here $h = 0.7$ has been
adopted
(\cite{TGB:holmberg},
\cite{McKay:dynam},
\cite{BS03},  \cite{TGB:dynam}).

\item Hosts must be at least 2.5 times brighter than any other galaxy
that is within a projected radius of $r_p < 700$~kpc and a relative
radial velocity difference of $|dv| < 1000$~km~sec$^{-1}$.
Satellites must be at least 6.25 times fainter than their host,
must be found within $r_p < 500$~kpc,
and the host--satellite
velocity difference must be $|dv| < 500$~km~sec$^{-1}$. Here 
$h = 0.7$ has been adopted
(\cite{TGB:holmberg}, \cite{SL}).
\end{enumerate}
Although the above criteria may seem lax or even somewhat arbitrary,
in the case of the first two sets of criteria, both the Milky Way
and M31 would be excluded from the sample of hosts.  That is, these
particular
selection criteria give rise to samples of unusually isolated host
galaxies.  In addition, both 
Prada et al.\ \cite{Prada} and Brainerd \cite{TGB:holmberg} adopted a number of
different selection criteria in their investigations of the satellites
of SDSS galaxies and concluded that there were no statistical differences
between results that were obtained with different selection criteria.
In other words, provided sufficiently ``reasonable''  criteria are adopted for
selecting isolated hosts and their satellites, the results of the
investigations are stable to modest differences in the details of those
selection criteria.

\begin{figure}
\centerline{
\scalebox{0.85}{%
\includegraphics{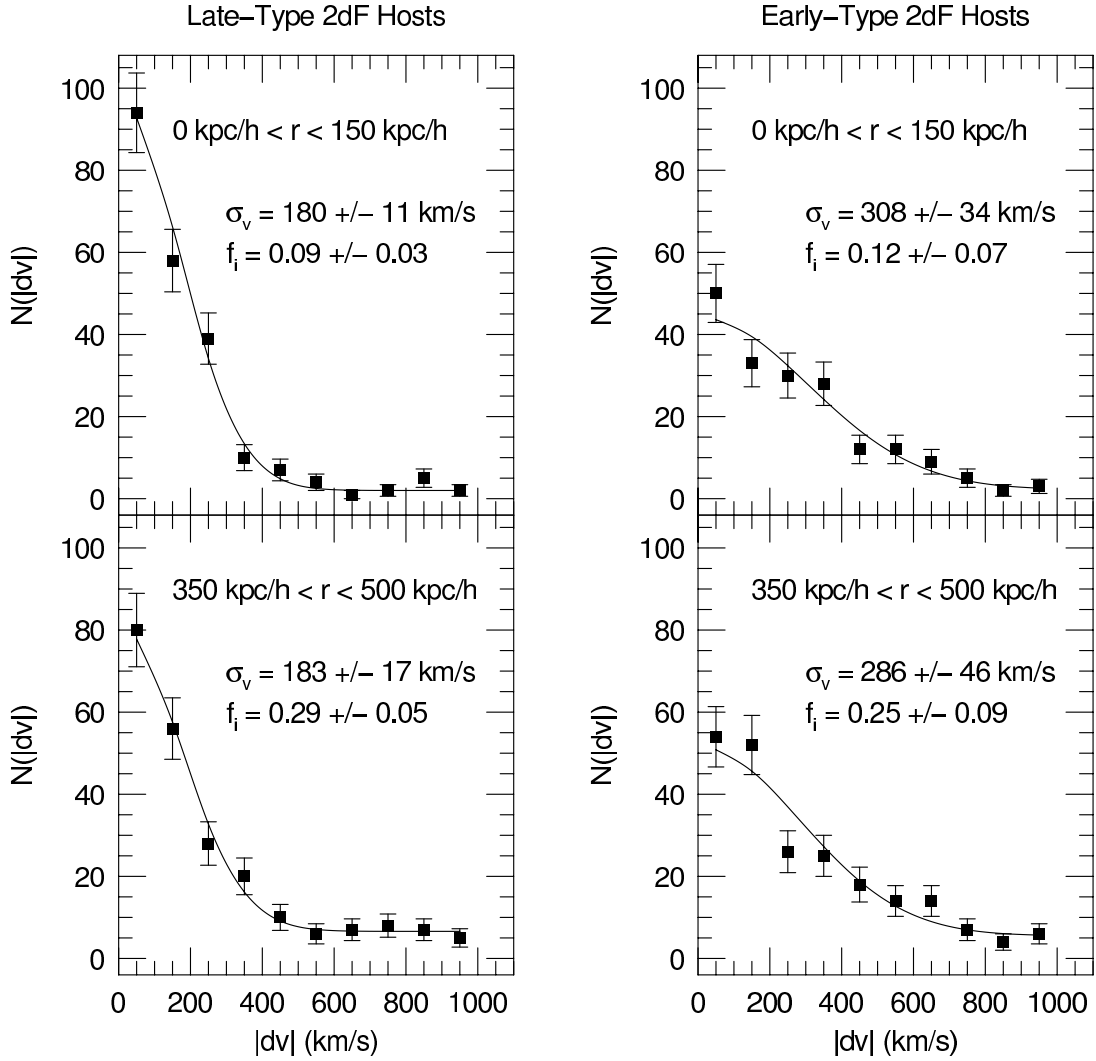}%
}
}
\caption{
Points with error bars show the
observed distribution of velocity differences, $N(|dv|)$, for 
a subset of host--satellite
systems in the 2dFGRS for which the host morphologies have been
visually classified.
Solid lines show the best--fitting
``Gaussian plus offset'' function, from which the velocity dispersion of
the satellites, $\sigma_v$, and the fraction of interlopers,
f$_{\rm i}$, is determined.
Left panels: late--type hosts.
Right panels: early--type hosts.  Top panels: satellites located close
to the host in projection on the sky.  Bottom panels: satellites located
far from the host in projection on the sky.  A substantially
larger value of $\sigma_v$ is obtained for the satellites of early--type hosts
than for the satellites of late--type hosts.  Note, too, that the fraction
of interlopers increases significantly with the projected radius, $r$, of
the satellites.
}
\label{dv:fit}
\end{figure}

No matter what selection criteria are adopted, however, there
will always be ``interlopers'' in the satellite data.  Interlopers
are galaxies that are falsely identified as satellites; that is, they
pass the formal selection criteria, but they are not, in fact, dynamically
associated with the host galaxy.  
The presence of interlopers will artificially inflate any measurement
of the velocity dispersion of genuine satellites, and recent
investigations of satellite
dynamics (\cite{Prada},
\cite{McKay:dynam}, \cite{BS03}, \cite{TGB:dynam})
have corrected for the effects of interlopers by modeling the 
distribution of host--satellite velocity differences
as the sum of a Gaussian distribution (due to the genuine
satellites) and a constant offset (due to the interlopers).  Prada et al.\
\cite{Prada} used numerical 
simulations to show that this is a sensible way in which to correct
for the effects of interlopers.  Moreover, both Brainerd \& Specian
\cite{BS03} and Prada et al.\ \cite{Prada} have pointed out that an
accurate determination of the velocity dispersion profile, $\sigmav (r_p)$,
for satellite galaxies depends on a proper determination of the 
interloper fraction as an explicit function of the projected radius.
That is, by purely geometrical effects, the interloper fraction is
necessarily an increasing function of $r_p$.  
An example
of fitting a ``Gaussian  plus offset'' to the distribution of velocity
differences for late--type galaxies and early--type galaxies  in the
2dFGRS is
shown in Figure~\ref{dv:fit}.  One can clearly see from this
figure that the velocity
dispersion of the satellites is a function of the morphology of the
host galaxy (being larger for early--type hosts than late--type
hosts), and that the interloper fraction increases with projected 
radius.

The above ``Gaussian plus offset'' fit to the distribution of host--satellite
velocity differences accounts for the fact that the number of
interlopers is a function of projected radius, and it assumes {\it a
priori} that the number of interlopers at a given projected radius is constant
with $|dv|$.
Recently, however,
van den Bosch et al.\ \cite{vandenBosch:2004b} used simulations
of galaxy redshift surveys to investigate this and found a sharp
increase in the number of interlopers for small relative velocities.
van den Bosch et al.\ \cite{vandenBosch:2004b} note, however, that
the value of $\sigmav$ that is determined from a simple ``Gaussian
plus offset'' fit is not strongly affected by the fact that the
number of interlopers varies with $|dv|$.  This is because the
best--fitting value of $\sigmav$
is rather insensitive to the precise value of the interloper fraction.
Brainerd \cite{TGB:dynam} also finds that the number of interlopers
is larger for small values of $|dv|$ than it is for large values of
$|dv|$, but that the effect is not nearly as pronounced as found
by van den Bosch et al.\ \cite{vandenBosch:2004b}.  Given the size
of the error bars on the distribution of host--satellite velocity
differences in the current observational samples, then, it would appear
that the simple ``Gaussian plus offset'' fit to the distribution of
velocity differences is more than adequate to the task of estimating
$\sigmav (r_p)$.

\section{Theory: ``Universal'' (NFW) Halos vs.\ Isothermal Halos}

High--resolution CDM simulations have established the existence
of a ``universal'' density profile for dark matter halos which results
from generic dissipationless collapse
(e.g., \cite{NFW:1995}, \cite{NFW:1996},
\cite{NFW:1997}, \cite{Matthias:1998}, \cite{Thomas:1998}, 
\cite{Kravtsov:1997}, \cite{Tormen:1997}, \cite{Ghigna:1998}, 
\cite{Ben:1998}). 
This density profile fits objects that
span roughly 9 orders of magnitude in mass (ranging from the
masses of globular star 
clusters to the masses of large galaxy clusters) and applies
to physical scales
that are less than the ``virial'' radius, $r_{200}$.  
Conventionally, $r_{200}$
is defined to be the radius at which the spherically--averaged mass
density reaches 200 times the critical mass density (e.g., \cite{NFW:1995},
\cite{NFW:1996}, \cite{NFW:1997}).

Navarro, Frenk \& White \cite{NFW:1995}, \cite{NFW:1996}, \cite{NFW:1997}
showed that the universal density profile for dark 
matter halos was fitted well by a function of the form
\begin{equation}
\rho(r) =
\frac{\delta_{c}\rho_{c}}{\left(r/r_{s}\right)
\left(1+r/r_{s}\right)^{2}},
\label{rho:nfw}
\end{equation}
and halos having such a density profile are generally referred to
as ``NFW'' halos.
Here $\rho_{c} = \frac{3H^{2}(z)}{8\pi G}$ is the critical density 
of the universe
at the redshift, $z$, of the halo, $H(z)$ is Hubble's
parameter at that same redshift, and  $G$ is Newton's constant.
The scale radius $r_{s} \equiv r_{200}/c$ is a characteristic radius 
at which the density profile agrees with the isothermal profile
(i.e., $\rho(r) \propto r^{-2}$),
$c$ here is a dimensionless number known as the concentration parameter, and
\begin{equation}
\delta_{c}= \frac{200}{3}\frac{c^{3}}{\ln(1+c)-c/(1+c)}
\end{equation}
is a characteristic overdensity for the halo.

Formally, the above fitting function for the radial density profiles
of CDM halos converges to a steep, cuspy profile: $\rho(r) \propto
r^{-1}$.  The NFW fitting formula, however, was never intended to be
extrapolated to very small radii (i.e., radii smaller than the practical
resolution limits of the simulations) and much fuss has been made over
whether observed galaxies actually
show such cuspy inner density profiles (e.g., 
\cite{Ben:1994}, \cite{FP:1994}, \cite{McGaugh:1998}, \cite{Debattista:1998},
\cite{Ben:1999},
\cite{vandenBosch:2000}, \cite{Beto1}, \cite{Beto2}).  
More recent numerical work has
shown that the density profiles of CDM halos do not, in
fact, converge to a well--defined asymptotic
inner slope (e.g., \cite{Power:2003}, \cite{Stoehr:2003}
\cite{Julio:2004}, \cite{Huss}), and it has
become increasingly clear that fair and direct comparisons of simulated galaxies
with observed galaxies on very small physical scales is an extremely
challenging thing to do (e.g., \cite{Rhee}, \cite{Joel}).

Weak lensing and satellite dynamics do not have the ability to provide
any information whatsoever on the cuspiness (or lack thereof) in the
central regions of
galaxies.  Instead, both are governed by the large--scale properties
of the halos (i.e., the regime in which the NFW profile is known to
be an excellent description of the density profiles of CDM halos) and,
at least in principle, both have the potential to discriminate between
NFW halos and simpler singular isothermal sphere halos. 

The radial density profile of a singular isothermal sphere halo is
given by
\begin{equation}
\rho(r) = \frac{\sigma_v^2}{2\pi G r^2}
\label{rho:sis}
\end{equation}
(e.g., \cite{Binney:Tremaine}), where $\sigmav$ is the velocity dispersion.
The isothermal sphere is characterized by the single parameter $\sigmav$,
which is constant as a function of radius.  
A key prediction for NFW halos, however,
is that the radial velocity dispersion
will have a strong dependence upon the radius and this, of course,
is inconsistent with
the constant value of the velocity dispersion
that characterizes singular isothermal spheres.  Specifically,
on sufficiently small scales $\sigma_r (r)$ should {\it increase}
with radius, and on large scales $\sigma_r (r)$ should {\it decrease}
with radius.  
Hoeft, M\"ucket \& Gottl\"ober \cite{Hoeft:2004}  have shown that
the radial velocity dispersion of NFW halos can be fitted by a
function of the gravitational potential, $\Phi(r)$, of the form:
\begin{equation}
\sigma_r (r) = \left[
a \left(\Phi_{\rm out} - \Phi(r) \right)
\left( \frac{\Phi(r)}{\Phi_{\rm out}} \right)^\kappa
\right]^{1/2} .
\label{nfw:sigmav}
\end{equation}
Note that $\sigma_r(r)$ above
is not the ``line of sight'' velocity dispersion, since
$r$ is a true 3--dimensional radius in 
eqn.\ (\ref{nfw:sigmav}).  The parameters $a$ and $\kappa$
have values of $a = 0.29 \pm 0.04$ and $\kappa = 0.41 \pm 0.03$, and
$\Phi_{\rm out}$ is the outer potential of the halo.
Therefore, we expect the dynamics within an NFW halo to 
differ fundamentally from the dynamics within an isothermal 
sphere halo.

In the case of weak lensing, NFW halos give rise to a distortion in
the images of distant galaxies that differs somewhat from the distortion
that 
would be yielded by an isothermal sphere halo (e.g., \cite{Matthias:1996},
\cite{WB2000}).  The radial dependence of the shear
for the isothermal sphere is given by:
\begin{equation}
\gamma_{\rm sis}(r_p) = \frac{2\pi}{r_p} \left( \frac{\sigmav}{c}  \right)^2 
\frac{D_{ls} D_l}{D_s}
\label{gamma:sis}
\end{equation}
(e.g., \cite{dasBuch}).  Here $c$ is the velocity of light and
$D_s$, $D_l$, and $D_{ls}$ are again angular diameter distances.
In the case of NFW halos,
the radial dependence
of the shear is given by:
\begin{equation}
\gamma_{\rm nfw}(x) = \left\{ \begin{array}{ll}
\frac{r_{s}\delta_{c}\rho_{c}}{\Sigma_{c}}
g_{<}(x) & \mbox{$\left(x < 1\right)$} \\
 & \\
\frac{r_{s}\delta_{c}\rho_{c}}{\Sigma_{c}}
\left[\frac{10}{3} + 4\ln\left(\frac{1}{2}\right)\right] & \mbox{$\left(x = 1\right)$} \\
 & \\
\frac{r_{s}\delta_{c}\rho_{c}}{\Sigma_{c}}
g_{>}(x) & \mbox{$\left(x > 1\right)$}
\end{array}
\right.
\label{gamma:nfw}
\end{equation}
where $x \equiv r_p/r_s$, $\Sigma_c$ is the critical mass density for
gravitational lensing given by eqn.~(\ref{Sigmac}), and 
the functions $g_{<,>}(x)$ 
are explicitly independent of the cosmology:
\begin{eqnarray}
\hspace{-0.4cm}g_{<}(x) & \hspace{-0.15cm}= \hspace{-0.15cm}& \frac{8{\rm arctanh}
\sqrt{\frac{1-x}{1+x}}}{x^{2}\sqrt{1-x^{2}}}
+ \frac{4}{x^{2}}\ln\left(\frac{x}{2}\right) - \frac{2}{\left(x^{2}-1\right)}+
\frac{4{\rm arctanh}\sqrt{\frac{1-x}{1+x}}}{\left(x^{2}-1\right)\left(1-x^{2}\right)^{1/2}}  \label{gless} \\
\hspace{-0.4cm}g_{>}(x) &\hspace{-0.15cm} =\hspace{-0.15cm} & \frac{8\arctan\sqrt{\frac{x-1}{1+x}}}{x^{2}\sqrt{x^{2}-1}}
\hspace{0.1cm}
+ \hspace{0.1cm}\frac{4}{x^{2}}\ln\left(\frac{x}{2}\right) - \frac{2}{\left(x^{2}-1\right)}
\hspace{0.1cm}+\hspace{0.1cm}
\frac{4\arctan\sqrt{\frac{x-1}{1+x}}}{\left(x^{2}-1\right)^{3/2}}.
\label{ggtr}
\end{eqnarray}
(e.g., \cite{Matthias:1996}, \cite{WB2000}). 

In the following sections I summarize the most recent attempts to study the
dark matter halos of field galaxies through satellite dynamics and weak
lensing, including attempts to distinguish between isothermal and NFW 
potentials on the basis of the velocity dispersion profile and on the weak
lensing shear.

\section{Observed Velocity Dispersion Profiles}

At best, galaxy--galaxy lensing and satellite dynamics have the potential
to constrain the dependence of the line of sight velocity
dispersion on the projected radius, $\sigmav (r_p)$.  
Determining $\sigmav (r_p)$ has proven to be quite a challenge to
galaxy--galaxy lensing studies, in large part because the shear profiles
of NFW lenses and isothermal sphere lenses are not dramatically different,
except on the very smallest ($r < r_s$) and very largest ($r > r_{\rm vir}$)
scales \cite{WB2000}.  To date, only one 
tentative measurement of $\sigmav (r_p)$ has been made from observations
of galaxy--galaxy lensing \cite{combo17:old}.  Kleinheinrich
et al.\ \cite{combo17:old}  modeled the lens galaxies in the COMBO--17 survey
as singular isothermal spheres with velocity dispersions that scaled
with luminosity as
\begin{equation}
\frac{\sigmav}{\sigmav^\ast} =
\left( \frac{L}{\Lstar} \right)^\eta
\label{TF}
\end{equation}
where $\sigmav^\ast$ is the line of sight
velocity dispersion of the halo of an
$\Lstar$ galaxy.  Kleinheinrich et al.\ \cite{combo17:old} fixed $\eta$ to be 0.35
and determined best-fitting values of $\sigmav^\ast$ for projected
radii in the range $20~h^{-1}~{\rm kpc} < r_p < r_{\rm max}$.  When
they considered all lenses in their sample, Kleinheinrich et al.\ 
\cite{combo17:old}
found $\sigmav^\ast \sim 139$~\kms for $r_{\rm max} = 50~h^{-1}$~kpc,
$\sigmav^\ast \sim 164$~\kms for $r_{\rm max} = 150~h^{-1}$~kpc,
and $\sigmav^\ast \sim 123$~\kms for $r_{\rm max} = 500~h^{-1}$~kpc. 
This suggests a velocity dispersion profile that rises at small
radii, reaches a maximum, then decreases at large radii.
However, the formal error bars on these measurements show that
all of these values of $\sigmav^\ast$ agree to 
within one to two standard deviations.
In addition, it should be kept in mind that each of these measurements of
$\sigmav^\ast$ is not independent (as they would be if
a differential measurement of $\sigmav^\ast (r_p)$ were made), so the
data points and their error bars are all correlated with one another.

Considerably stronger constraints on the dependence of the halo velocity
dispersion with projected radius have come from the most recent investigations
of the motions of satellites about host galaxies.  In particular, both
Prada et al.\ \cite{Prada} and Brainerd \cite{TGB:dynam} have measured
decreasing velocity dispersion profiles for the satellites of host
galaxies in the SDSS and 2dFGRS, respectively.  Although they used different
data sets and different host--satellite selection criteria, both
Prada et al.\ \cite{Prada} and Brainerd \cite{TGB:dynam} used the same
technique to make measurements of the velocity dispersion profiles.
That is, the distribution of velocity differences, $N(|dv|)$,
for satellites found within projected radii of $r_{\rm min} < r_p <
r_{\rm max}$ was modeled as a combination of a Gaussian and an offset
due to interlopers.  In both studies, the interloper fraction was determined
separately for each of the independent radial bins.

\begin{figure}
\centerline{
\scalebox{0.95}{%
\includegraphics{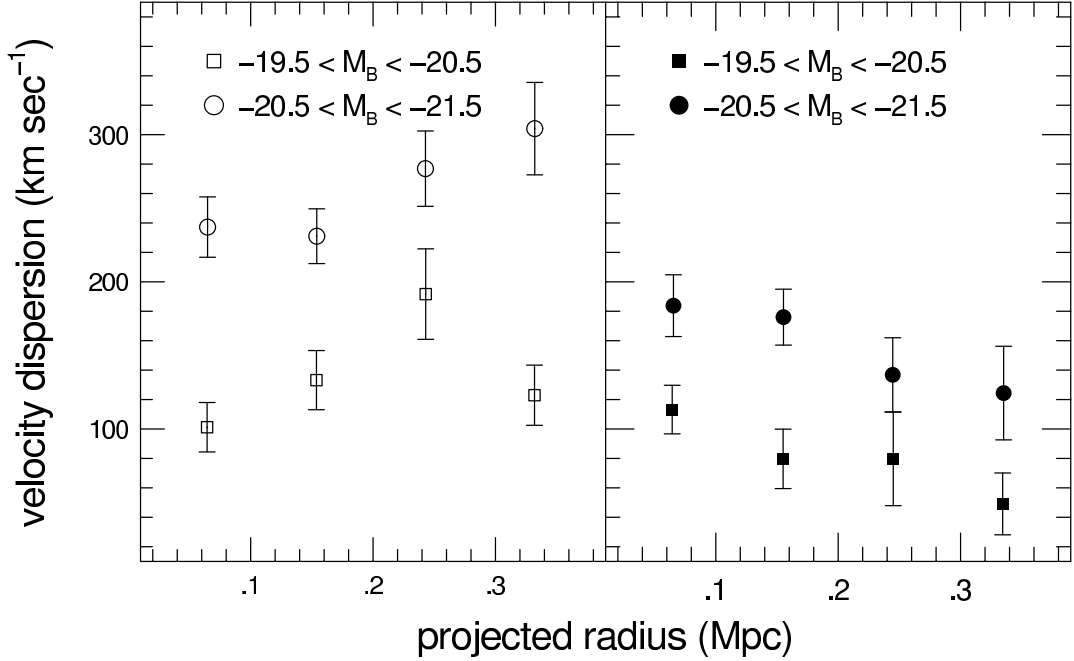}%
}
}
\caption{
Velocity dispersion profiles for satellites of SDSS host galaxies
\cite{Prada}.  Circles: host galaxies with $-20.5 < M_B < -21.5$, Squares:
host galaxies with $-19.5 < M_B < -20.5$.  Left panel: ``raw'' velocity 
dispersion profiles prior to correction for contamination by interlopers.
Right panel: velocity dispersion profiles after correction for contamination
by interlopers.  After correction for interlopers, $\sigmav (r_p)$ for
the satellites of the
fainter hosts is consistent with the expectations for an NFW
halo with $M_{200} = 1.5\times 10^{12} \Msun$, and $\sigmav (r_p)$ for
the satellites of 
the brighter hosts is consistent with the expectations for an
NFW halo with $M_{200} = 6\times 10^{12} \Msun$.
Here $h = 0.7$ has been adopted.
}
\label{Prada:sigmav}
\end{figure}

Prior to correcting for the contamination of interlopers, Prada et al.\
\cite{Prada} found a velocity dispersion profile, $\sigmav (r_p)$, that
increased with 
projected radius.
After the removal of the interlopers, however, Prada et al.\ \cite{Prada}
found
decreasing velocity dispersion profiles in both cases.  The corresponding
velocity dispersion profiles are shown in
Figure~\ref{Prada:sigmav}. Moreover, their
corrected velocity dispersion profiles were fitted well by the velocity
dispersion profiles of NFW halos with virial masses of
$1.5\times 10^{12} ~\Msun$ (hosts with absolute magnitudes $-19.5 < M_B < -20.5$)
and $6\times 10^{12} ~\Msun$ 
(hosts with absolute magnitudes $-20.5 < M_B < -21.5$).  
Since Prada et al.\ \cite{Prada} adopted a value of $h = 0.7$ and since the
absolute magnitude of an $L^\ast$ galaxy is $M_B^\ast \sim -19.5$, these
results suggest that the virial mass of the halo of an $L^\ast$ galaxy is
$\lo 10 \times 10^{11} h^{-1} \Msun$.

\begin{figure}
\centerline{
\scalebox{0.90}{%
\includegraphics{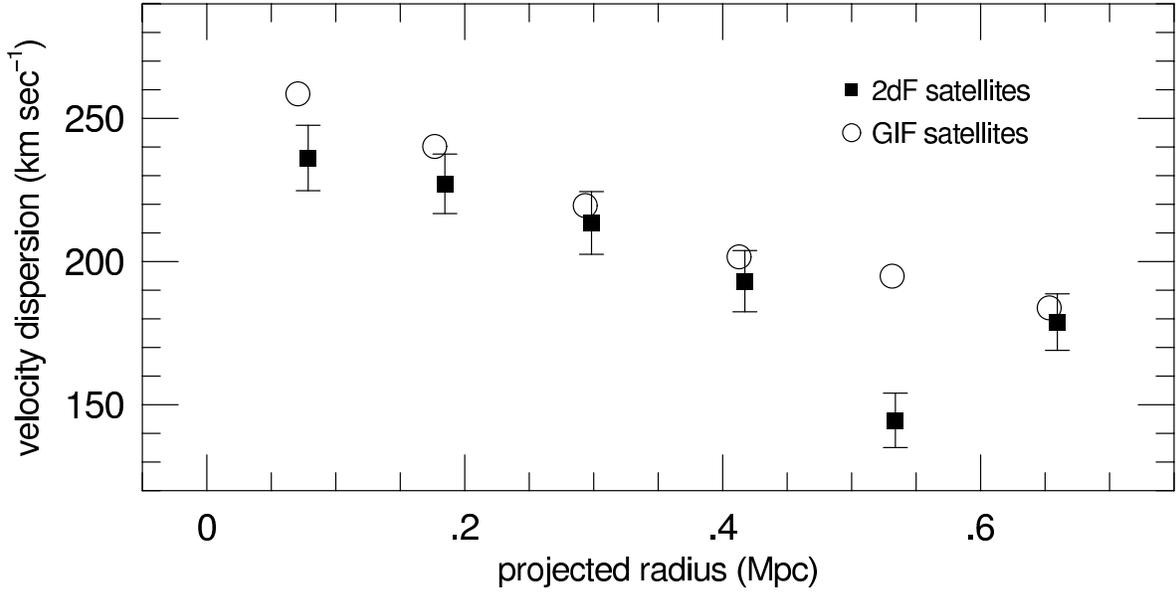}%
}
}
\caption{
Velocity dispersion profiles for satellites in the final data release
of the 2dFGRS and the flat, $\Lambda$--dominated GIF simulation 
\cite{TGB:dynam}.
Here $h = 0.7$ has been adopted.
}
\label{2dF:sigmav}
\end{figure}

Brainerd \cite{TGB:dynam} selected hosts and satellites from the
final data release of the 2dFGRS using criteria identical
to those of Sample 3 in Prada et al.\ \cite{Prada}.  In addition,
she used these same criteria to select hosts and satellites from
the present epoch galaxy catalogs of the flat, $\Lambda$--dominated
the GIF simulation \cite{GIF}.  This is a publicly--available simulation
which includes semi--analytic galaxy formation in a CDM universe.
Brainerd \cite{TGB:dynam} restricted her analysis to hosts with
luminosities in the range
$0.5~ \Lstar \le L \le 5.5~ \Lstar$, and found a roughly similar number
of hosts and satellites in both the 2dFGRS (1345 hosts, 2475 satellites)
and the GIF simulation ($\sim 1200$ hosts, $\sim 4100$ satellites, 
depending upon the viewing angle).  Like Prada et al.\ \cite{Prada},
Brainerd \cite{TGB:dynam} obtained a decreasing velocity dispersion
profile for the satellites of the 2dFGRS galaxies once the effects of
interlopers were removed.  In addition, excellent agreement between
$\sigmav (r_p)$ for the 2dFGRS galaxies and $\sigmav (r_p)$ for 
the GIF galaxies was found, showing consistency between the motions
of satellites in the 2dFGRS and the expectations of a $\Lambda$--dominated
CDM universe.  See Figure \ref{2dF:sigmav}.

\begin{figure}
\centerline{
\scalebox{0.95}{%
\includegraphics{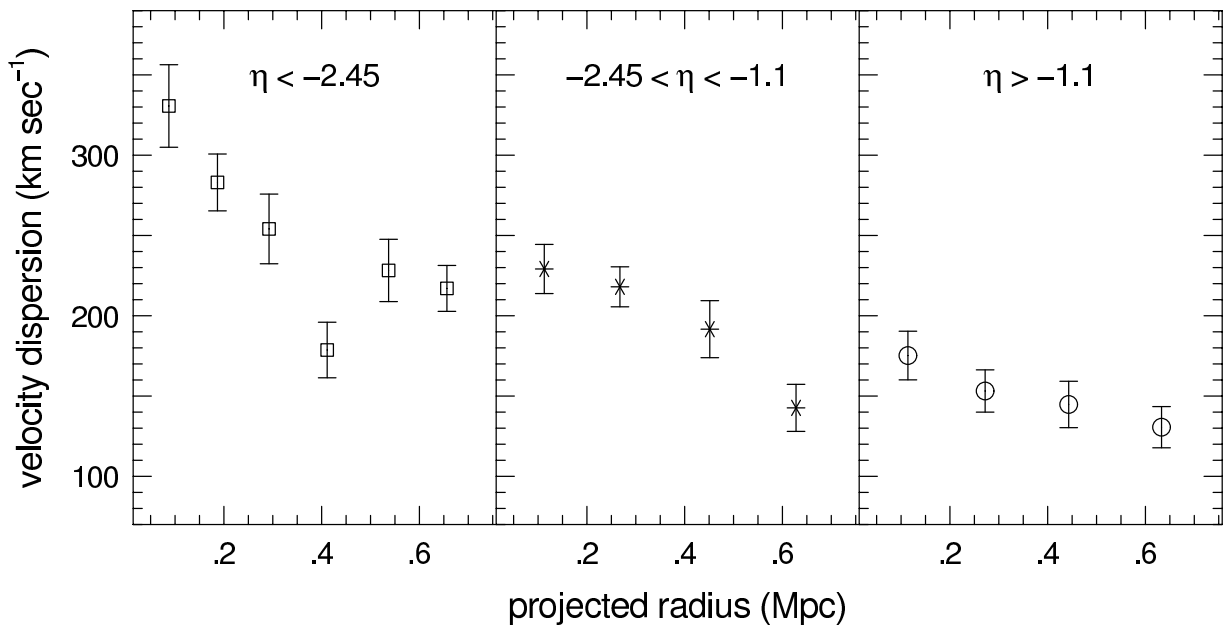}%
}
}
\caption{
Velocity dispersion profiles for satellites in the final data release
of the 2dFGRS as a function of the host spectral parameter, $\eta$ 
\cite{TGB:dynam}.
The morphology of the hosts is expected to be roughly
E/S0 in the left panel, Sa in the middle panel, and Sb/Scd in the
right panel. 
The median luminosities of the subsamples in each of the panels is
somewhat different, but the difference is too small to account for the
differences in the velocity dispersion profiles.
Here $h = 0.7$ has been adopted.
}
\label{2dF:svspec}
\end{figure}

Further, Brainerd \cite{TGB:dynam} divided her sample of 2dFGRS
host galaxies into thirds based upon the spectral index parameter,
$\eta$ \cite{Madgwick}, and computed the dependence of the velocity dispersion
profile on host spectral type.  The subsamples corresponded to hosts 
which are expected to have morphologies that are approximately:
(i) E/S0, (ii) Sa, and (iii) Sb/Scd.  The median luminosities of 
the hosts in the subsamples were all fairly similar:
(i) $2.64~L_{b_J}^\ast$, (ii) $2.25~L_{b_J}^\ast$, and (iii)
$2.11~L_{b_J}^\ast$.  The velocity dispersion profiles of all three
samples decreased with radius and, moreover,  
$\sigmav (r_p)$ was found to have a much higher
amplitude and steeper decline for the satellites of early--type
hosts than it did for the satellites of
late--type hosts.  See Figure \ref{2dF:svspec}.  Although there is
some difference in the median luminosities of the hosts in the
subsamples, the difference is too small to have a significant effect on the
velocity dispersion profiles.  Therefore, the results of Brainerd 
\cite{TGB:dynam} seem to indicate that early--type galaxies
have deeper potential wells (and hence
more massive halos) than late--type galaxies. 

Previous 
work on the dependence of $\sigmav$ with projected
radius using SDSS galaxies \cite{McKay:dynam} and
2dFGRS galaxies \cite{TGB:dynam} concluded that
$\sigmav (r_p)$ was consistent with an isothermal profile; i.e., 
$\sigmav (r_p) = {\rm constant}$.
In both of these investigations, the hosts and satellites were selected in a
manner that was identical to that of Sample 3 in Prada et al.\
\cite{Prada}.  In both previous analyses, however, the number of 
hosts and satellites
was significantly smaller than the more recent studies, and the
formal error bars were correspondingly larger. In addition,
the original analysis of SDSS host--satellite systems \cite{McKay:dynam}
neglected to account for the fact that the interloper fraction
increases with radius, which would have biased measurements of
$\sigmav$ at large $r_p$ towards values which are higher than
the actual satellite velocity dispersion at those radii.

Even more recently, Conroy et al.\ \cite{DEEP2} used 
satellites of $z \sim 0.8$ host galaxies 
in the DEEP2 survey to investigate $\sigmav (r_p)$.  DEEP2 
(Deep Extragalactic Evolutionary Probe 2) is
being carried out with the DEIMOS spectrograph at the Keck--II
telescope, and will ultimately collect spectra of $\sim 60,000$
galaxies with redshifts of $0.7 \lo z \lo 1.4$ to a limiting
magnitude of $R_{AB} = 24.1$ \cite{Marc}.  Unfortunately, the
survey is still far from complete and only 61 isolated host
galaxies (having a total of 75 satellites) were found in the current
DEEP2 data.  Because of this, the errors on $\sigmav (r_p)$
are large, and formally $\sigmav (r_p)$ for the DEEP2 galaxies
is fitted well by a constant value:
$\sigmav (110~h^{-1}~{\rm kpc}) = 162^{+44}_{-30}$~\kms,
$\sigmav (230~h^{-1}~{\rm kpc}) = 136^{+26}_{-20}$~\kms,
$\sigmav (320~h^{-1}~{\rm kpc}) = 155^{+55}_{-38}$~\kms.
Therefore, isothermal halos for the DEEP2 galaxies cannot 
be ruled out at the moment.  Conroy et al.\ \cite{DEEP2} show,
however, that their velocity dispersion measurements are consistent
with expectations for NFW halos with virial masses in the
range $3.5\times 10^{12}~h^{-1}~\Msun \le M_{200} \le
8.0\times 10^{12}~h^{-1}~\Msun$.  This is in good 
general agreement 
with the results of Prada et al.\ \cite{Prada}, especially
considering that the DEEP2 hosts are of order one magnitude
brighter than the SDSS hosts (i.e., the virial mass implied for
the halos of the brightest galaxies in the SDSS sample is
$\sim 4\times 10^{12} h^{-1} \Msun$).  At the moment, however, the 
DEEP2 data are too severely limited by small number statistics
to place strong constraints on the nature of the dark matter
halos of galaxies with redshifts of order unity.

\section{Halo Masses and Galaxy Mass--to--Light Ratios}

Although it ought to be straightforward and even easy to compare
the halo masses and galaxy mass--to--light ratios that are obtained
from different studies, in practice it is rather like comparing
persimmons to tomatoes; i.e., they are vaguely similar on the inside
and outside, but they are definitely not interchangeable.  
The fundamental problem is that it is simply not
possible to measure the ``total'' mass of a galaxy halo (since it
is not possible to say where such a halo ``ends'') and, hence, all
halo masses are simply masses that are contained within some 
physical radius of the center of the halo.  Along those same lines,
and given that velocity dispersion profiles of NFW halos decrease
with radius, if one wants to compare the results of two investigations
which have measured a velocity dispersion averaged over some large
scale, it is important that those scales be {\it identical}.  That
is, suppose a single measurement of $\sigma_v$ is made by averaging over
scales $r < 100~h^{-1}$~kpc in one study and a single measurement of
$\sigma_v$ is made by averaging over scales $r < 200~h^{-1}$~kpc
in another.  If the second measurement of $\sigma_v$ is lower than
the first by some significant amount, that does not necessarily mean
that the values are in disagreement.  They would be in disagreement
if both halos were isothermal spheres, but if the halos
are NFW objects, then it is only to be expected that the second measurement
would be lower than the first.

A more subtle problem is the definition of the ``virial radius'' in the
context of NFW halos.  While
$r_{200}$ was originally proposed as the radius at which the interior
mass density is 200 times the critical mass density 
(e.g., \cite{NFW:1995}, \cite{NFW:1996}, \cite{NFW:1997}), it is not
at all uncommon to find that investigators who have fit NFW models to
their data have defined the virial radius as the radius at which the
interior mass density is 200 times the mean mass density of the universe.
Therefore, what is meant by a ``virial mass'' in the context of an
NFW fit to data can (and does) vary from investigation to investigation,
and a certain amount of care has to be taken when comparing such results.
Despite the difficulties of comparing the conclusions of different
studies, I will forge ahead because it is becoming clear that a consistent
picture really is emerging on the topic of the masses of the halos of field
galaxies, and their corresponding mass--to--light ratios.  The weak
lensing studies yield results that are by and large consistent with
each other, and the dynamical studies seem to be in general agreement
with the trends in the weak lensing data: the halos have masses that
are consistent with expectations for galaxy--sized halos in CDM, 
and there are real,
physical differences between halos surrounding (i)
early--type and late--type
galaxies and (ii) high--luminosity and low--luminosity galaxies.

\subsection{$M$ and $M/L$ from Galaxy--Galaxy Lensing}

\begin{figure}
\centerline{
\scalebox{1.05}{%
\includegraphics{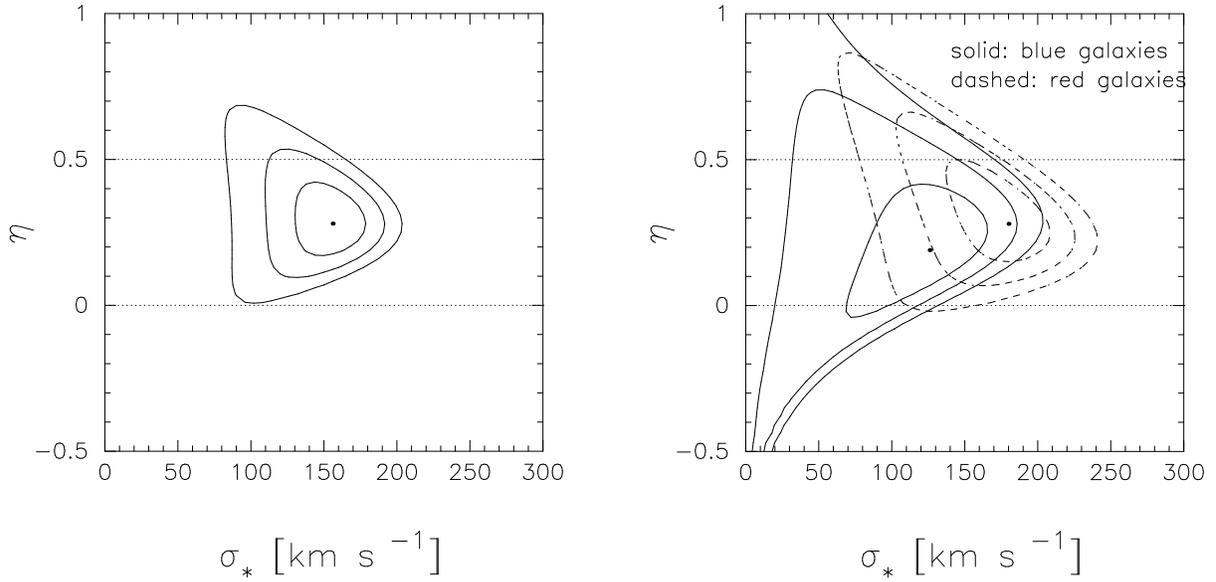}%
}
}
\caption{
Isothermal sphere models for the galaxy--galaxy data from
 COMBO--17 \cite{combo17:new}.  Joint constraints
(1$\sigma$, 2$\sigma$, and 3$\sigma$) on the velocity
dispersion, $\sigma_v^\ast$, of the halos of $L^\ast$ galaxies and
the index of the Tully--Fisher/Faber--Jackson relation, $\eta$.
Here the weak lensing signal has been averaged over scales
$r_p \lo 150~h^{-1}$~kpc.
Left panels: all lenses, $\sigma_v^\ast = 156^{+18}_{-18}$~\kms,
$\eta = 0.28^{+0.12}_{-0.09}$.  Right panels:
red lenses (2579 galaxies, $\sigma^\ast_{v,~{\rm red}} = 180^{+24}_{-30}$~\kms)
Left panels: blue lenses (9898 galaxies, $\sigma^\ast_{v,~{\rm blue}} =
126^{+30}_{-36}$~\kms).
Figure kindly provided by Martina Kleinheinrich.
}
\label{Martina:sis}
\end{figure}

In the case of galaxy--galaxy lensing, it is not possible at the moment
to discriminate between shear profiles that are caused by NFW versus
isothermal galaxy halos.  Therefore, investigators will often choose one or
the other to constrain the properties of the halos that are producing
the lensing signal.  In the case of isothermal sphere halos, the 
velocity dispersions of the lens galaxies used to model the observed
signal are often chosen to scale as in eqn.~(\ref{TF}) above,
$(\sigmav/ \sigmav^\ast) = 
( L/L^\ast )^\eta$,
where again $\sigmav$ is the velocity dispersion of a halo that contains
a galaxy of luminosity $L$, and $\sigmav^\ast$ is the velocity dispersion
of the halo of
an $L^\ast$ galaxy.  Hoekstra et al.\ \cite{Henk:2004} used this
approach with their RCS data, as did Kleinheinrich et al.\ 
\cite{combo17:new} with their COMBO--17 data.  When all lenses and 
sources were used in the investigations, and when the lensing signal
was averaged over an identical scale ($r \lo 350~h^{-1}$~kpc), 
both the RCS and COMBO--17 results
are in very good agreement with each other.  In particular, Hoekstra
et al.\ \cite{Henk:2004} find $\sigmav^\ast = 136 \pm 8$~\kms for
an adopted value of $\eta = 0.3$, and Kleinheinrich et al.\ 
\cite{combo17:new} find $\sigmav^\ast = 138^{+18}_{-24}$ and
$\eta = 0.34^{+0.18}_{-0.12}$.  Further, Kleinheinrich et al.\ 
\cite{combo17:new} find
that there are clear differences in the halos surrounding ``blue''
galaxies (rest frame colors of $(U-V) \le 1.15 - 0.31z - 
0.08[M_V - 5 \log h + 20]$) and those surrounding
``red'' galaxies (the remainder
of the sample).  That is, the red COMBO--17 lens galaxies have a higher
velocity dispersion than the blue COMBO--17 lens 
galaxies, but both have a similar value of
the index $\eta$ above.  See Figure~\ref{Martina:sis}.

\begin{figure}
\centerline{
\scalebox{0.55}{%
\includegraphics{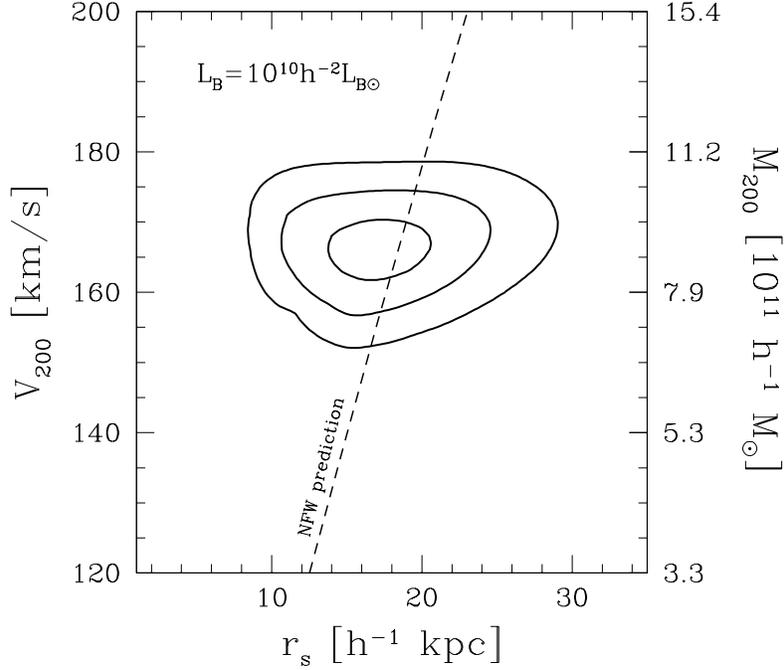}%
}
}
\caption{
Constraints on the circular velocity at $r = r_{200}$ and the scale
radius, $r_s$, for lenses in the RCS that have been modeled as
having NFW--type halos \cite{Henk:2004}.  Formally, the best--fitting
values of the circular velocity, scale radius and virial mass are:
$V_{200} = 162 \pm 8$~\kms, $r_s = 16.2^{+3.6}_{-2.9}~h^{-1}$~kpc,
and $M_{200} = 8.4\pm 1.1 \times 10^{11}~h^{-1}~\Msun$.  Here $r_{200}$
is defined as the radius at which the mean interior mass density of
the halo is equal to $200 \rho_c$.
The dashed line shows the predictions of the NFW theory, in which
$V_{200}$ and $r_{200}$ are not independent parameters.
Figure kindly provided by Henk Hoekstra.
}
\label{Henk:nfw}
\end{figure}

\begin{figure}
\centerline{
\scalebox{1.05}{%
\includegraphics{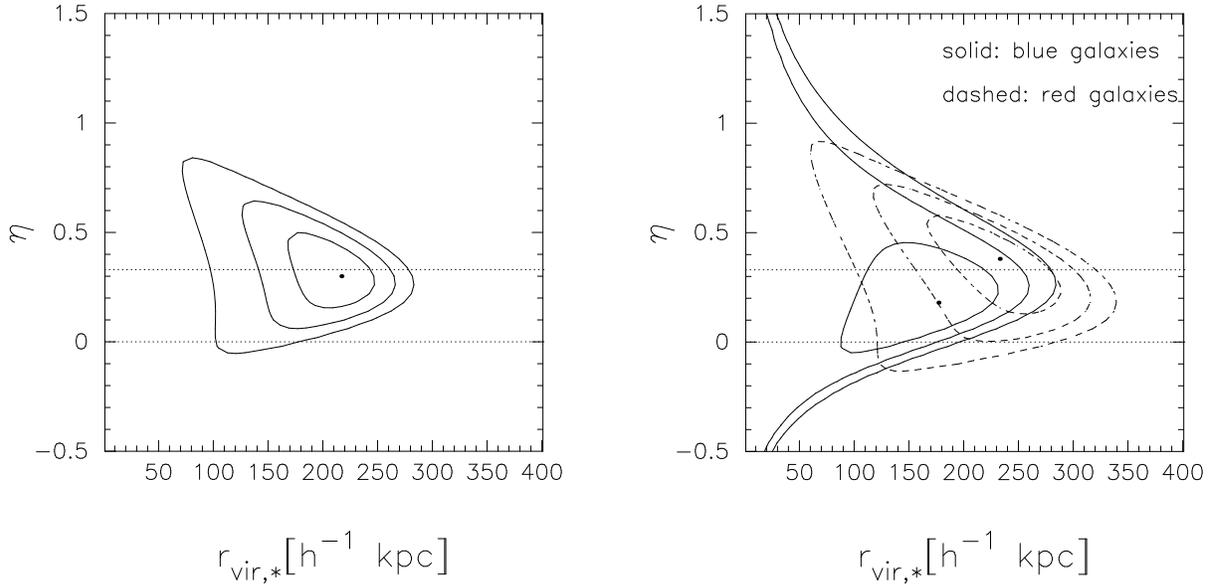}%
}
}
\caption{
NFW halo models of the galaxy--galaxy lensing data from COMBO--17
\cite{combo17:new}.   Joint constraints (1$\sigma$, 2$\sigma$, and
3$\sigma$) on $\eta$ and the virial radii of the halos of $L^\ast$
galaxies are shown.  Left panel: all lenses, $\eta = 0.30^{+0.16}_{-0.12}$,
$r_{\rm vir}^\ast = 217^{+24}_{-32}~h^{-1}$~kpc.  Right panel: red 
lenses (2579 galaxies, $\eta = 0.38^{+0.16}_{-0.20}$,
$r_{\rm vir}^\ast = 233^{+48}_{-48}~h^{-1}$~kpc) versus blue lenses 
(9898 galaxies, $\eta = 0.18^{+0.16}_{-0.16}$,
$r_{\rm vir}^\ast = 177^{+40}_{-56}~h^{-1}$~kpc).
Figure kindly provided by Martina Kleinheinrich.
}
\label{Martina:nfw}
\end{figure}

In addition,
Guzik \& Seljak \cite{Guzik:Seljak}, Hoekstra et al.\
\cite{Henk:2004}, and Kleinheinrich et al.\ \cite{combo17:new} have
all used NFW halos to model their lens galaxies, and all find very
reasonable fits to their lensing signals.  Further, the derived values
of the NFW virial masses of the halos of $L^\ast$ galaxies are in quite
good agreement amongst these studies when they are determined in 
similar band passes (e.g., $r$) and with identical definitions of the virial
radius \cite{combo17:new}: $M_{\rm vir}^\ast = 8.96 \pm 1.59 \times
10^{11}~h^{-1}~\Msun$ \cite{Guzik:Seljak}, $M_{\rm vir}^\ast = 8.4 \pm 0.7
\times 10^{11}~h^{-1}~\Msun$ \cite{Henk:2004}, and $M_{\rm vir}^\ast = 
7.8^{+3.5}_{-2.7} \times 10^{11}~h^{-1}~\Msun$ \cite{combo17:new}.
These are also in remarkably good agreement with the virial mass implied
for the halos of $L^\ast$ galaxies by the dynamical analysis of
Prada et al.\ \cite{Prada} (e.g., $M_{\rm vir}^\ast \sim 10 \times 10^{11}
h^{-1} \Msun$).
Shown in Figure~\ref{Henk:nfw}
are 1$\sigma$, 2$\sigma$ and 3$\sigma$ confidence limits on a joint--parameter
fit of the circular velocity at $r_{200}$, $V_{200}$, and scale radius,
$r_s$, for the lenses in the RCS data \cite{Henk:2004}.  Note that in
the analysis of the RCS data, $V_{200}$ and $r_s$ were allowed to vary
freely, while, to within some scatter, these parameters
are strongly correlated in the NFW
theory (i.e., the NFW model is in essence specified by a single
parameter).  The dashed line in Figure~\ref{Henk:nfw} therefore shows the
prediction for a strict adherence to the NFW theory (i.e., 
$V_{200}$ and $r_s$ are correlated appropriately), and the fact that
the theoretical NFW line passes so well through the contours gives
a certain amount of confidence that the NFW model is a very good fit
to the data.
Kleinheinrich et al.\ \cite{combo17:new} find good fits of the NFW 
model to their data and, moreover, find that both the virial radii
of the halos and the parameter $\eta$ are
dependent upon the rest frame colors of the galaxies, with red galaxies
having a somewhat larger virial radius (and, hence, larger virial
mass) than blue galaxies.    See Figure~\ref{Martina:nfw}. 
Here $\eta$ is defined not as in eqn.~(\ref{TF}), since the
velocity dispersion is a function of projected radius in the NFW model,
but rather it is defined as:
\begin{equation}
\frac{r_{\rm vir}}{r_{\rm vir}^\ast} = 
\left( \frac{L}{L^\ast} \right)^\eta,
\end{equation}
in analogy to the Tully--Fisher and Faber--Jackson relations (see
\cite{combo17:new}).  In this
case, $r_{\rm vir}^\ast$ is the virial radius of the halo of an
$L^\ast$ galaxy, defined at 200 times the mean mass density of the universe.
The variation of
$\eta$ with galaxy color and its implications for the mass--to--light
ratios of the galaxies will be discussed below.

A particularly detailed study of the masses of lensing galaxies as
a function of their color was carried out by Guzik \&
Seljak \cite{Guzik:Seljak} for $\sim 3.5\times 10^4$ lenses and
$\sim 3.6\times 10^6$ sources in the SDSS.  All of the lens galaxies
have spectroscopic redshifts in this case, and all of the halos were
modeled as NFW objects in the context of the ``halo model''.
In all 5 of the SDSS band passes, Guzik \& Seljak \cite{Guzik:Seljak}
find that the virial masses of $L^\ast$ ellipticals exceed those
of $L^\ast$ spirals though, unsurprisingly, the amount by which 
the masses of the ellipticals exceeds those of the spirals is a
strong function of the band pass.  In the redder bands, the masses
of the ellipticals exceed those of the spirals by a factor of
$\sim 2$ to $\sim 2.5$, while in $g'$ the difference is a factor
of $\sim 6$ and in $u'$ the difference is close to an order of 
magnitude.  Although it is difficult to make direct comparisons between
the two studies (because of the differing definitions of the virial
radius and the different definitions of the subsamples of galaxies),
there is good general agreement between the results of Guzik \&
Seljak \cite{Guzik:Seljak} and Kleinheinrich et al.\ \cite{combo17:new}:
when the galaxy--galaxy lensing signal is detected
red band passes (e.g., $R$, $r'$) and the lenses are modeled as
NFW objects, the virial masses of red/early--type galaxies exceed those
of blue/late--type galaxies by a factor of order 2.

In addition to the halos of early--type lenses having more mass
than those of late--type lenses, the weak lensing
work of Sheldon et al.\ \cite{Sheldon}
indicates that, again, in all 5 SDSS band passes, the projected excess
surface mass density increases with the luminosity of the lens.
Sheldon et al.\ \cite{Sheldon} separated their $\sim 1.27\times 10^5$ lenses
into 3 magnitude bins (high, middle, and low luminosity), and the 
magnitude cuts differ for the different band passes.  (See Table~2 of
Sheldon et al.\ \cite{Sheldon} for a complete list of the magnitude
cuts as a function of band pass.) In the case of the $r'$ data, the
``high'' luminosity galaxies have a mean absolute magnitude of -22.5,
the ``middle'' luminosity galaxies have a mean absolute magnitude
of -21.9, and the ``low'' luminosity galaxies have a mean absolute
magnitude of -20.5.  These mean
luminosities correspond roughly to $4.5 \Lstar$ (``high''),
$2.7 \Lstar$ (``middle'') and $0.8 \Lstar$ (``low'') in the $r'$ band.
In all cases, $\Delta \Sigma (r_p)$ for the ``high'' luminosity galaxies
exceeds that of the ``medium'' and ``low'' luminosity galaxies, and for
$r_p \lo 1~h^{-1}$~Mpc, the difference corresponds to an approximately
constant multiplicative factor.  Specifically at $r_p \sim 100~h^{-1}$~Mpc,
however, $\Delta \Sigma$ for the high luminosity lenses in Sheldon et al.\
\cite{Sheldon} exceeds that for the low luminosity lenses by
a factors of $\sim 3$ in $u'$, $\sim 5$ in $g'$, $\sim 5$ in $r'$,
$\sim 7$ in $i'$, and $\sim 7$ in $z'$ (e.g., Figure~14 of Sheldon
et al.\ \cite{Sheldon}).  Similar trends (i.e., higher projected excess
surface mass density for more luminous lenses) were found by Seljak et al.\
\cite{Uros} in their galaxy--galaxy lensing analysis of SDSS data.

Lastly, although there is reasonable agreement regarding the relative
increase in mass for the halos of early--type lens galaxies versus
late--type lens galaxies at fixed luminosity (i.e., $L^\ast$), there
is some disagreement over the dependence of the mass--to--light ratio
on the luminosity of the host.  Specifically, in their redder bands
Guzik \& Seljak \cite{Guzik:Seljak} find that the mass--to--light
ratio goes as $M/L \propto L^{0.4 \pm 0.2}$ for $L > \Lstar$, suggestive
of a mass--to--light ratio that increases with luminosity.  Kleinheinrich
et al.\ \cite{combo17:new}, however, find that $M/L$ for their sample
of lenses is more consistent with a constant
value: $M/L \propto L^{-0.10^{+0.48}_{-0.36}}$.  Both Guzik \& Seljak
\cite{Guzik:Seljak} and Kleinheinrich et al.\ \cite{combo17:new} agree,
however,
that the mass--to--light ratio of red/early--type $L^\ast$ lens galaxies
exceeds that of blue/late--type $L^\ast$ lens galaxies by a factor of
$\sim 2$ to $\sim 2.5$ in the redder bands.

\subsection{$M$ and $M/L$ from Satellite Dynamics}

In the 1990's, Zaritsky et al.\ \cite{ZSFW}  and
Zaritsky \& White \cite{Zaritsky:White} used the velocity differences between a
small number of isolated spiral galaxies and their satellites to show that the
halos of the spirals were massive and extended to large radii:
$M(150~h^{-1}~{\rm kpc}) \sim 1$ to $2\times 10^{12} h^{-1} \Msun$.  
Moreover, Zaritsky et al.\ \cite{ZSFW} found a somewhat curious result:
the velocity difference between their 115 satellites and 69 hosts was
independent of the inclination corrected H-I line width of the host and
was, therefore, independent of the luminosity of the host (through, e.g.,
the Tully--Fisher relation).  At fixed large radius, then, this would
imply that $M/L$ for the spiral hosts decreased as $M/L \propto L^{-1}$.

More recent investigations of halo masses and corresponding mass--to--light
ratios from satellite dynamics have led to rather a large assortment
of conclusions.  McKay et al.\ \cite{McKay:dynam} and Brainerd \& Specian
\cite{BS03} used the dynamics of the satellites of SDSS galaxies and
2dFGRS galaxies, respectively, to constrain the dynamical masses of the
halos of the host galaxies interior to a radius of $r = 260~h^{-1}$~kpc.
Both used an isothermal mass estimator of the form
\begin{equation}
M_{260}^{\rm dyn} = \frac{2.1~r~\sigmav^2}{G},
\label{mdynam:iso}
\end{equation}
where $\sigmav$ is the line--of--sight velocity dispersion.  Both felt
this assumption was justified because both found that their velocity 
dispersion profiles were consistent with a constant value.  In the case
of McKay et al.\ \cite{McKay:dynam}, however, no correction for an 
increasing number of interlopers with projected radius was made and this
may have led to an incorrect conclusion that $\sigmav(r_p)$ was independent
of $r_p$.  In the case of Brainerd \& Specian \cite{BS03}, the increasing
number of interlopers at large $r_p$ was taken into account, but only
galaxies from the 100k data release of the 2dFGRS were used (i.e., roughly
half as many galaxies as in the full data release), and although
$\sigma_v (r_p)$ was consistent with a constant value in their data,
the later analysis by Brainerd \cite{TGB:dynam} showed that this was
simply due to the rather large error bars in Brainerd \& Specian 
\cite{BS03}.  This being the case, the mass--to--light ratios published
by these two studies are suspect at some level, but it is unclear at
the moment just how suspect they may actually be.  That is, while it is 
true that the velocity dispersion profile of NFW halos decreases with
radius, the fall--off in $\sigmav (r_p)$
is not particularly sharp and it is not obvious
how badly isothermal mass estimates of the form in 
eqn.\ (\ref{mdynam:iso}), which are based on an average value of
$\sigmav$, will compare to proper NFW mass estimates.

Formally, McKay et al.\ \cite{McKay:dynam} found that in all 5 SDSS
band passes, $M_{260}^{\rm dyn}/L$ was roughly constant for $L > L^\ast$,
and that the value of $M_{260}^{\rm dyn}/L$ was a strong function of 
the band pass (being systematically higher in the blue bands than in
the red bands).  Brainerd \& Specian \cite{BS03} found that for 
$L \go 2 L^\ast$, $M_{260}^{\rm dyn}/L$ was a constant for dynamical
analyses that included (i) all 809 hosts in their sample and (ii) 159 hosts
that had been visually classified as early--type (E/S0).  However, much
like the results of Zartisky et al.\ \cite{ZSFW}, Brainerd \& Specian
\cite{BS03} found that $M_{260}^{\rm dyn}/L$ 
decreased as $M_{260}^{\rm dyn}/L \propto L^{-1}$ for 243 hosts that had been
visually classified as spirals.  This latter result remains puzzling,
and is certainly in need of further investigation with larger data sets.

\begin{figure}
\centerline{
\scalebox{0.90}{%
\includegraphics{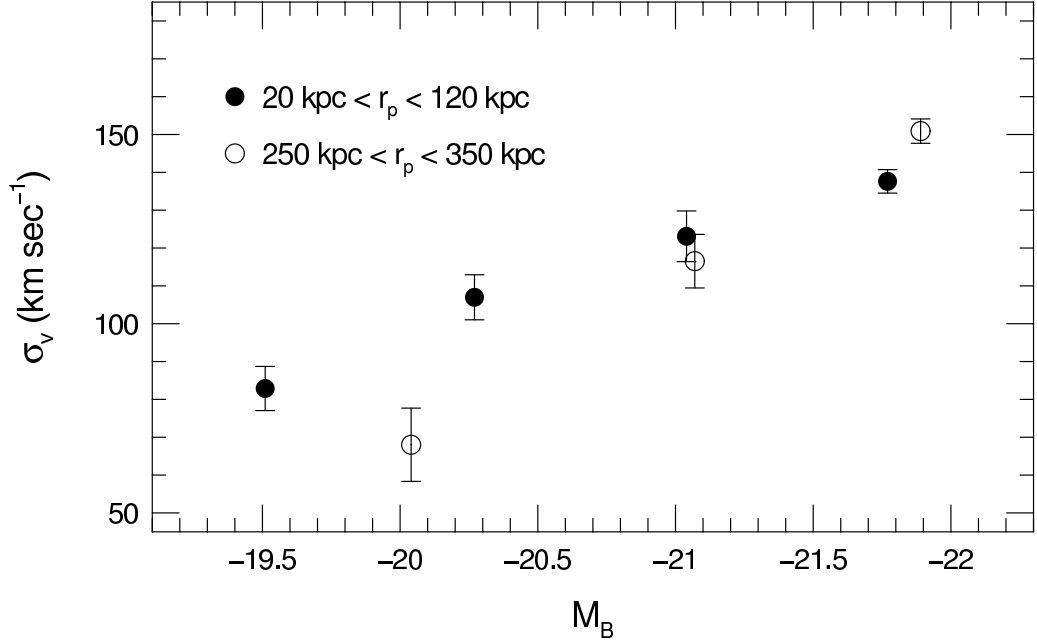}%
}
}
\caption{
Dependence of satellite velocity dispersion on host absolute
magnitude for SDSS galaxies \cite{Prada}.  Filled circles:
$\sigmav$ computed using satellites with $20~{\rm kpc} \le r_p
\le 120~{\rm kpc}$.  Open circles: $\sigmav$ computed using 
satellites with $250~{\rm kpc} \le r_p \le 350~{\rm kpc}$.
For small projected radii the velocity dispersion scales
as  $\sigmav \propto L^{0.3}$, in good
agreement with the local B--band Tully--Fisher relationship.  For
large projected radii $\sigmav \propto L^{0.5}$.
Here $h = 0.7$ has been adopted.
}
\label{Prada:TF}
\end{figure}

In their analysis of the dynamics of the satellites of SDSS host
galaxies, Prada et al.\ \cite{Prada} found
that the velocity dispersion of the satellites scaled with host luminosity
as $\sigmav \propto L^{0.3}$ (i.e., in good agreement with the local
$B$--band Tully--Fisher relationship \cite{Verheijen}) for satellites
with projected radii $r_p < 120$~kpc.   (Recall, too, that in this
study $\sigmav (r_p)$ was specifically corrected for the increase in
interlopers at large $r_p$.)  In addition, Prada et al.\ \cite{Prada}
found that for satellites at large projected radius,
$250~{\rm kpc} < r_p <
350~{\rm kpc}$, the velocity dispersion scaled with luminosity
as $\sigmav \propto L^{0.5}$ (i.e., steeper than expected from the
Tully--Fisher relation).  See Figure~\ref{Prada:TF}.

\begin{figure}
\centerline{
\scalebox{1.0}{%
\includegraphics{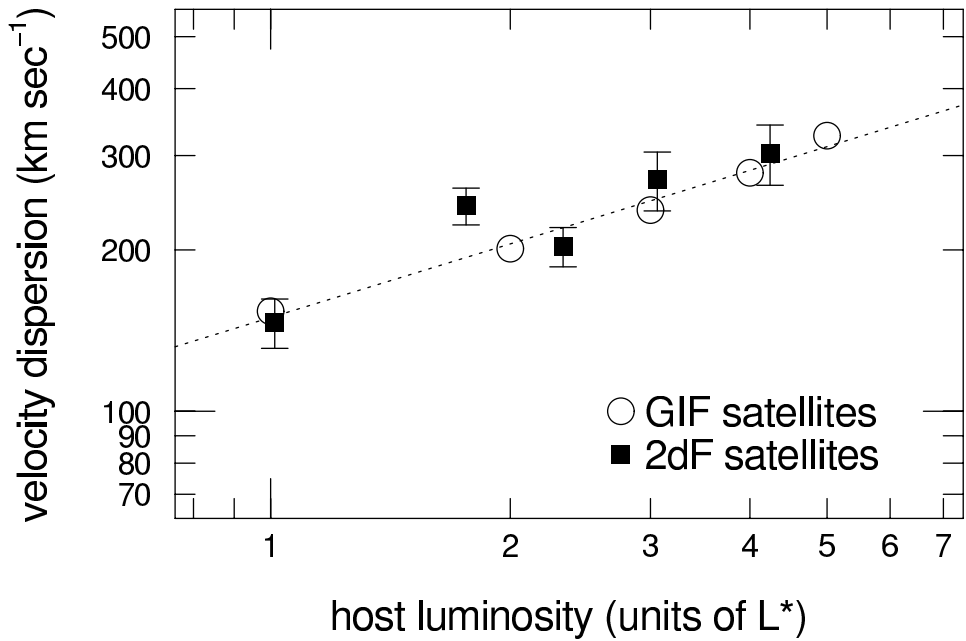}%
}
}
\caption{
Dependence of satellite velocity dispersion on host luminosity
for satellites with projected radii
$r_p \le 120$~kpc in both the 2dFGRS and the flat,
$\Lambda$--dominated GIF simulation \cite{TGB:dynam}.
Dotted line shows $\sigmav \propto L^{0.45}$.
}
\label{2dF:TF}
\end{figure}

Similar to Prada et al.\ \cite{Prada}, Brainerd \cite{TGB:dynam}
also computed the dependence of the small--scale velocity dispersion
of satellites on host luminosity.  See Figure~\ref{2dF:TF}.  Like
Prada et al.\ \cite{Prada}, Brainerd \cite{TGB:dynam} corrected for the fact that
the interloper fraction is an increasing function of projected radius and
overall, she found excellent agreement
between the velocity dispersions of satellites
with projected radii $r_p \le 120$~kpc in the 2dFGRS and
GIF simulations.  The velocity dispersions of the 2dFGRS satellites
were, however, seen to scale with host luminosity as $\sigmav \propto
L_{b_J}^{0.45 \pm 0.10}$, which is only marginally consistent
with the results of Prada et al.\ \cite{Prada} and the local
$B$-band Tully--Fisher relationship.

Prada et al.\ \cite{Prada} have shown (e.g., their Figure~12) that the dependence
of the line of sight velocity dispersion on the virial
mass of NFW halos scales as
$\sigmav \propto M_{\rm vir}^{0.38}$ for the case that $\sigmav$ is
computed as an average over scales $20~{\rm kpc} \lo r_p \lo 100~{\rm kpc}$,
and that $\sigmav \propto M_{\rm vir}^{0.50}$ for the case that
$\sigmav$ is computed at $r_p \sim 350$~kpc.  Combining this with their
results for the dependence of $\sigmav$ on $L$ at different scales leads
to the conclusion that on scales $r_p \lo 120$~kpc, 
$M_{\rm vir}/L \propto L^{-0.2}$ while on scales $r_p \sim 300$~kpc,
$M_{\rm vir}/L$ is a constant.  Similarly, if the halos of the 2dFGRS
galaxies studied by Brainerd \cite{TGB:dynam} are assumed to be NFW 
objects, the implication is that $M_{\rm vir}/L \propto L^{0.2^{+0.3}_{-0.1}}$
for the 2dFGRS hosts (again, computed on scales $r_p \lo 120$~kpc). 

While it certainly cannot be said that there is a consensus from weak
lensing and satellite dynamics
as to the exact dependence of the galaxy mass--to--light ratio
on $L$, it does seem to be the
case that all of these studies point towards a dependence of $M_{\rm vir}/L$
on $L$ that is, at most, rather weak.  That is, with the notable exception
of the Brainerd \& Specian \cite{BS03} result for late--type galaxies,
all of the recent determinations of $M/L$ for $L \go L*$ find that,
to within 2$\sigma$,
$M/L$ is independent of $L$.
In addition, when the weak lenses and host galaxies are each
modeled as NFW objects, a fairly consistent value of the average
virial mass of the halos of $L^\ast$ galaxies is found: 
$\sim (8 - 10) \times 10^{11}~h^{-1}~\Msun$.
Further, it seems to be clear that 
both weak lensing and satellite dynamics indicate that the
masses of the halos of early--type galaxies are larger than that of
late--type galaxies, and that at fixed luminosity
the mass--to--light ratios of early--type galaxies are larger than
those of late--type galaxies.  

\section{Non--spherical Halos}

Although the simple 
isothermal sphere can reproduce the flatness of the rotation
curves of the disks of spiral galaxies at large radii, 
there are both observational
and theoretical arguments in favor of halos which are flattened, rather
than spherical.  Direct observational evidence for halo
flattening that has come from studies of
individual galaxies is somewhat scarce, however, owing to
the fact that there are relatively few galaxies for which the shape
of the halo potential can be probed directly.
Nevertheless, the evidence for flattened halos of individual
galaxies is
diverse and includes such observations as
the dynamics of polar ring galaxies, the
geometry of X-ray isophotes, the flaring of HI gas in spirals, the
evolution of gaseous warps, and the kinematics of Population II stars
in our own Galaxy.  In particular, studies of disk systems which probe
distances of order 15~kpc from the galactic planes suggest that the
ratio of shortest to longest principle axes of the halos is
$c/a = 0.5 \pm 0.2$ (see, e.g., the comprehensive review by Sackett
 \cite{Penny} and references therein).
Studies of a number of strong lens galaxies have also suggested that the
mass distributions of the lenses are not precisely spherical.
For example, Maller et al.\ \cite{Ari} 
found that, provided the disk mass is small compared to the halo mass,
the halo of the spiral galaxy which lenses the quasar B1600+434 is
consistent with $c/a = 0.53$.
In addition, the 17 strong lens systems studied by Keeton, Kochanek \&
Falco \cite{Keeton:1998}
showed some preference for flattened mass distributions, although
extremely flattened (i.e., ``disky'') mass distributions were ruled out.
Finally, a recent analysis of the luminous halos of 1047 edge--on
disk galaxies in the SDSS suggests that the old stellar populations of
these galaxies consist of moderately flattened spheroids with 
axis ratios of $c/a \sim 0.6$ \cite{Zibetti}.

On the theoretical side, high--resolution simulations
of dissipationless CDM models consistently produce
markedly non--spherical  galaxy halos
with a mean projected ellipticity of
$\epsilon \sim 0.3$
(see, e.g., \cite{DC:1991}, \cite{MSW}).
It is known, however, that the dark matter will react to the 
condensation of baryons during galaxy formation (e.g., \cite{Blumenthal})
and that the resulting increase in the central density leads to a more
spherical shape than if dissipation were not considered (e.g., 
\cite{Dubinski:1994}).
Recent simulations performed by Kazantzidis et al.\ \cite{Kazantzidis}
show that on scales $r << r_{\rm vir}$, the effects
of gas cooling cause a 
substantial circularization of the mass density profile,
leading to a
projected ellipticity of $\epsilon \sim 0.4$ to 0.5 in the
inner regions of the galaxy.
However, on scales $r \sim r_{\rm vir}$ Kazantzidis 
et al.\ \cite{Kazantzidis}
find that the projected ellipticity is $\epsilon \sim 0.3$.
Since both the weak lensing shear and satellite dynamics are determined
primarily by the large--scale mass distribution of the halos, the roundness
of the mass distribution on small scales due
to gas cooling should not have a dramatic effect.
From a theoretical standpoint, therefore, it is
not at all unreasonable to expect that 
galaxy--galaxy lensing and satellite dynamics should reflect a
significant flattening of the halos.

\subsection{Evidence for Flattened Halos from Galaxy--Galaxy Lensing}

Unlike a spherically--symmetric lens for which the gravitational lensing
shear is isotropic about the lens center, the shear due to an elliptical
lens is anisotropic about the lens center.
Specifically, at a given angular distance, $\theta$,
from an elliptical lens,
source galaxies which are located closer to the major axis
of the mass distribution of the lens
will experience greater shear than sources which
are located closer to the minor axis (e.g., \cite{dasBuch}).
Noting this,
Natarajan \& Refregier \cite{Priya:Fridge} and Brainerd \& Wright \cite{BW}
modeled
the dark matter halos of field galaxies as infinite
singular isothermal ellipsoids
and made rough estimates of the sizes of observational data sets which would
be required to detect ``anisotropic'' galaxy--galaxy lensing and, hence,
to constrain the net flattening of the halo population.  Both studies concluded
that, if the mean flattening of the halos is of order 0.3, then only
a relatively modest amount of imaging data would be necessary to
observe the effects of halo flattening on the weak lensing signal.

In estimating the amount of data that would be required to detect anisotropic
galaxy--galaxy lensing,
both Natarajan \& Refregier \cite{Priya:Fridge}
 and Brainerd \& Wright \cite{BW} made the
simplifying assumption that each distant source galaxy is lensed by only
one foreground galaxy.  However, for a somewhat deep imaging survey
($I_{\rm lim} \sim 23$), the simulations of galaxy--galaxy
lensing performed by Brainerd, Blandford
\& Smail \cite{BBS} indicated that most of the galaxies with
magnitudes in the range
$22 \lo I \lo 23$ would
 have been lensed at a comparable level by two or
more foreground galaxies.
In a realistic data set, these
multiple weak deflections might significantly affect the signal--to--noise
that could be achieved when attempting to detect anisotropic galaxy--galaxy
lensing.  This motivated Wright \& Brainerd \cite{WB2002} to carry out detailed
Monte Carlo simulations of galaxy--galaxy lensing by flattened halos,
including the effects of multiple weak deflections
on the final images of distant galaxies. 

Wright \& Brainerd \cite{WB2002} showed that multiple weak deflections
create systematic effects which could hinder
observational efforts to use weak lensing to constrain the projected
shapes of the dark matter halos of field galaxies.  They modeled the dark
matter halos of lens galaxies as truncated singular isothermal ellipsoids, and
for an observational data set in which the galaxies had
magnitudes in the range
$19 \lo I \lo 23$, they found that
multiple deflections resulted in strong correlations
between the post--lensing image shapes of most foreground--background
pairs of galaxies.  Imposing a simple redshift cut during
the analysis of the data set,
$z_l < 0.5$ and $z_s > 0.5$, was sufficient to reduce the correlation
between the final images of lenses and sources to the point that
the expected anisotropy in the weak lensing signal was  detectable via
a straightforward average.  Wright \& 
Brainerd \cite{WB2002} concluded that previous theoretical calculations
of weak lensing due to flattened 
halos had considerably underestimated the
sizes of the observational
data sets which would be required to detect this effect.  In particular,
for a multi--color survey in which the galaxies had apparent magnitudes of
$19 \lo I \lo 23$ and the imaging quality was modest,
Wright \& Brainerd \cite{WB2002}
found that a 4$\sigma$ detection could be obtained with a survey
area of order 22~sq.~deg.,
provided
photometric redshift estimates were made for the galaxies, the
typical error in $z_{\rm phot}$ was $\lo 0.1$, and only
source galaxies with azimuthal coordinates that were within $\pm 20^\circ$
of the lens symmetry axes were used in the data analysis.

To date, only one intrepid team of investigators has claimed a detection
of flattened halos from observations of galaxy--galaxy lensing.  In
their analysis of the RCS galaxy--galaxy lensing signal Hoekstra,
Yee \& Gladders \cite{Henk:2004}
took the approach of modeling the lens galaxies
as having halos with ellipticities that scaled linearly with the
ellipticity of the image of the lens:
$\epsilon_{\rm halo} = f \epsilon_{\rm light}$.
Further, they assumed that the major axis of the lens image was aligned
with the major axis of the halo in projection on the sky.  This is a
sensible assumption provided the majority of the lenses are relaxed
systems, and it is justified at least partially by the observations
of Kochanek \cite{CSK} who found that the major axes of the mass and
light of strong lens galaxies were aligned to within $\sim 10^\circ$
in projection on the sky.

\begin{figure}
\centerline{
\scalebox{0.45}{%
\includegraphics{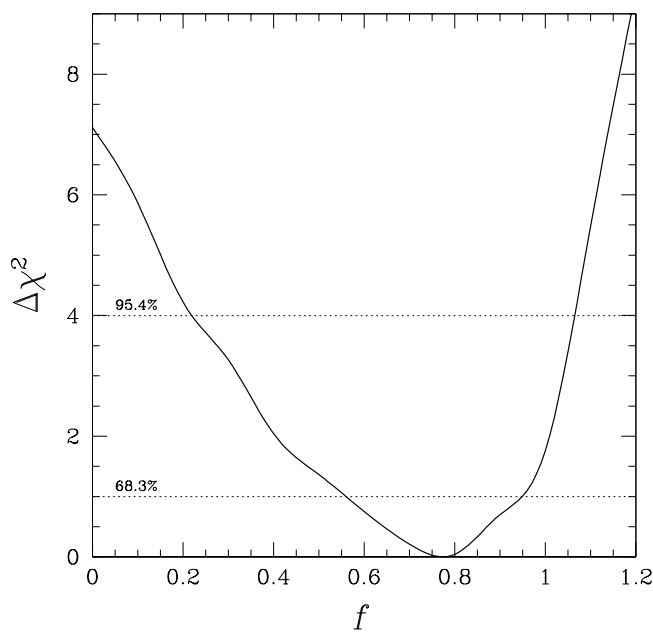}%
}
}
\caption{
Confidence bounds with which spherical halos can be rejected on the
basis of galaxy--galaxy lensing in the RCS \cite{Henk:2004}.
Halos of lens galaxies were modeled as having ellipticities of
$\epsilon_{\rm halo} = f 
\epsilon_{\rm light}$ and the principle axes of the halo mass were
assumed to be aligned with the symmetry axes of the lens images
in projection on the sky.  Round halos, $f = 0$, are excluded at
the 99.5\% confidence level.
Figure kindly provided by Henk Hoekstra.
}
\label{Henk:flat}
\end{figure}

Hoekstra, Yee \& Gladders \cite{Henk:2004} performed
a maximum
likelihood analysis 
and concluded that spherical halos
(i.e., $f = 0$) could be ruled out at the 99.5\% confidence level 
on the basis of their weak lensing signal (see Figure~\ref{Henk:flat}).  
Formally, Hoekstra, Yee \&
Gladders \cite{Henk:2004} found $f = 0.77^{+0.18}_{-0.21}$.  
Since the mean ellipticity
of the lens images in their study was $\left< \epsilon_{\rm light} \right> =
0.414$, this implies a mean halo ellipticity of
$\left< \epsilon_{\rm halo} \right> = 0.33^{+0.07}_{-0.09}$ and
a projected axis ratio of $c/a = 0.67^{+0.09}_{-0.07}$.  This is in
excellent agreement with the expectations for CDM halos, as well
as previous observational constraints on halo flattening obtained
on large physical scales (see, e.g., \cite{Penny}).
While it may yet be a bit premature to call this result a ``definitive''
measurement of the flattening of field galaxy halos, it is certainly
impressive and the statistics will only improve as weak lensing
surveys become larger.

\subsection{Evidence for Flattened Halos from Satellite Galaxies}

In the case of substantially flattened halos of host galaxies, one
would naively expect that satellite galaxies would show a somewhat
anisotropic distribution about the host.  That is, barring possible 
effects due to infall rates and orbital decay, one would expect 
the satellites to have some preference for being located near to
the major axis of the host's halo.  Until very recently, however, such
an observation had not been confidently
made and, moreover, a preference for
clustering of satellite galaxies along the {\it minor} axes of
host galaxies has been reported at a statistically significant level
by a handful of authors (\cite{ZSFW}, \cite{SL}, \cite{Holmberg}).
The apparent alignment of satellite galaxies
with the minor axes of the host galaxies is often referred to as
the Holmberg effect and  
in the naive picture of satellite orbits in flattened potentials,
observations of the Holmberg effect
lead to the uncomfortable conclusion that not only is
the halo mass flattened, but it is also
anti--aligned with the luminous regions of the galaxy.

While one is tempted to dismiss the minor axis clustering of satellites
observed by Zaritsky et al.\ \cite{ZSFW} and Holmberg \cite{Holmberg}
as being due to some combination of selection biases and very small sample
sizes, it is not easy to use this argument for the results of
Sales \& Lambas \cite{SL}.  
In their study, Sales \& Lambas \cite{SL} selected
hosts and satellites from the final data release of
the 2dFGRS, with a resulting sample
size of 1498 hosts and 3079 satellites.  The satellites were constrained
to be within projected radii $r_p \le 500$~kpc of their host and
to be within a velocity difference $|dv| < 500$~\kms.  Further, host
images were required to have eccentricities of at least 0.1 in order
that the orientation of their major axes be well--determined.   When Sales \&
Lambas \cite{SL} searched their entire
sample for anisotropies in the distribution
of satellites about 2dFGRS hosts, their results were consistent with
an isotropic distribution.  However, when they
restricted their sample to only hosts
and satellites whose radial velocities differed by
$|dv| < 160$~\kms, an apparently strong detection of the Holmberg 
effect (i.e., minor axis clustering of the satellites) was found.

\begin{figure}
\centerline{
\scalebox{0.90}{%
\includegraphics{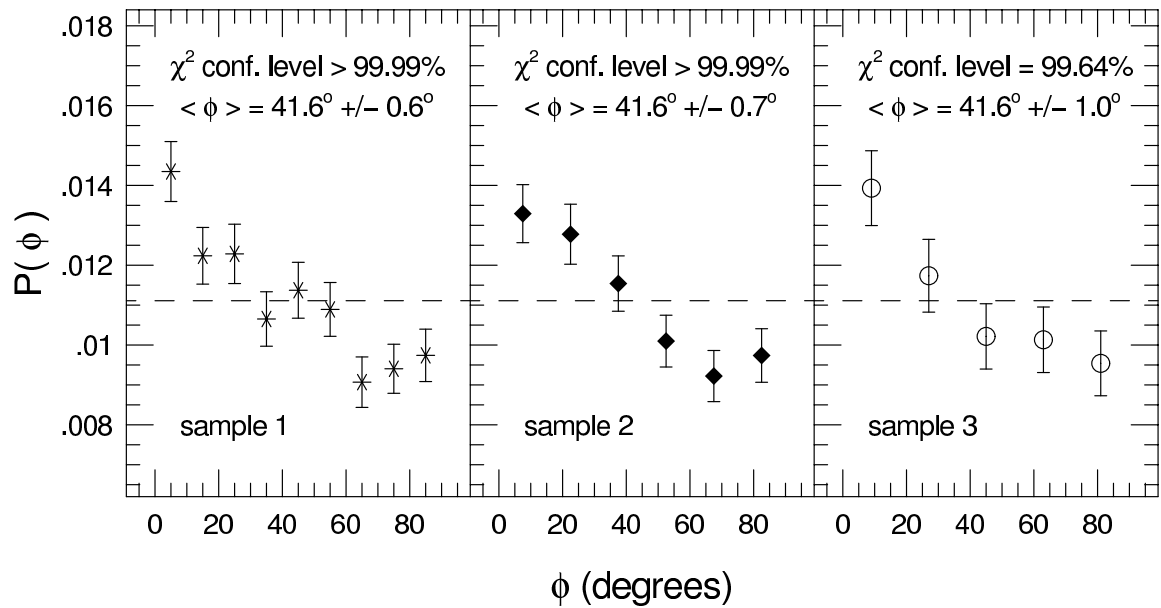}%
}
}
\caption{
Normalized probability distribution of the location
of satellite galaxies relative to the major axes of host galaxies
in the second data release
of the SDSS \cite{TGB:holmberg}.
Dashed line shows the expectation for an isotropic distribution.
Formal confidence levels at which isotropic distributions
can be rejected via $\chi^2$ tests are shown in each panel.
Also shown is $\phibar$, the mean value
of the angle between the major axis of the
host galaxy and the direction vector that connects the centroids
of the host and satellite.
}
\label{holmberg1}
\end{figure}

More recently, Brainerd \cite{TGB:holmberg} investigated the distribution
of satellites about hosts in the second data release of the SDSS.
She selected her samples using three different criteria: (1) the
criteria used by Sales \& Lambas \cite{SL} in their investigation of
the Holmberg effect for 2dFGRS galaxies, (2) the criteria used
by McKay et al.\ \cite{McKay:dynam} and Brainerd \& Specian \cite{BS03} in their 
analyses of satellite dynamics in the SDSS and 2dFGRS, respectively,
and (3) the selection criteria used by Zartisky et al.\ \cite{ZSFW}
in their investigation of the Holmberg effect.  
In addition, Brainerd \cite{TGB:holmberg} restricted the analyses to hosts with
ellipticities $\epsilon \ge 0.2$ and satellites that were found within
a projected radius of 500~kpc.
The three selection 
criteria lead to samples of: (1) 1351 hosts and 2084 satellites,
(2) 948 hosts and 1294 satellites, and (3) 400 hosts and 658 satellites
respectively.

\begin{figure}
\centerline{
\scalebox{0.90}{%
\includegraphics{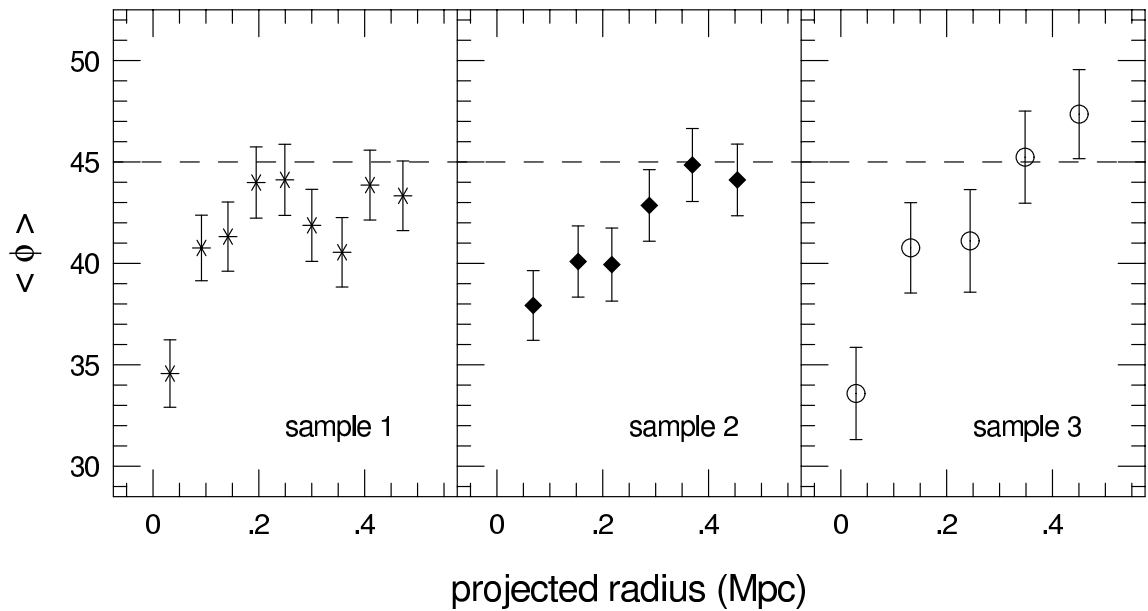}%
}
}
\caption{
Mean orientation of satellite galaxies with respect to the 
major axes of the hosts as a function of the projected radius for
galaxies in the second data release
of the SDSS \cite{TGB:holmberg}.
Dashed line shows the expectation for an isotropic distribution.
Here $h = 0.7$ has been adopted.
}
\label{holmberg2}
\end{figure}

In all three samples, Brainerd \cite{TGB:holmberg} found that the distribution of
satellites about their hosts was inconsistent with an isotropic distribution.
Formally, when a Kolmogorov--Smirnov test was applied to the
distribution of satellite locations, an
isotropic distribution was rejected at a confidence level of $> 99.99$\% 
for sample 1, $> 99.99$\% for sample 2,
and $99.89$\% for sample 3.  
Further, the mean angle between the major
axis of the host and the direction vector on the sky that connected the
centroids of the hosts and satellites was found to be
$\phibar = 41.6^\circ \pm 0.6^\circ$ for sample 1, $\phibar = 41.6^\circ \pm
0.7^\circ$ for sample 2, and $\phibar = 41.6^\circ \pm 1.0^\circ$ for
sample 3.  That is, a clear anisotropy in the distribution of satellites
about the hosts was seen, and the satellites showed a preference
for being aligned with the {\it major} axis of the host rather than
the minor axis (see Figure~\ref{holmberg1}).
In addition, Brainerd \cite{TGB:holmberg} investigated the dependence of $\phibar$
with projected radius on the sky and found that the majority of the
anisotropy arose on small scales ($\lo 200$~kpc) in all
three samples (see Figure~\ref{holmberg2}).  In other words, the
anisotropy was detected on physical scales that are comparable to the
expected virial radii of large, bright galaxies.  On scales much
larger than the expected virial radii of galaxy--sized halos
($r_p \sim 400$~kpc to 500~kpc), the distribution of satellites about
the SDSS hosts was consistent with an isotropic distribution at the
1$\sigma$ level.

Aside from the Brainerd \cite{TGB:holmberg} claim of ``planar'' (rather than
``polar'') alignment of satellites with the symmetry axes of their
hosts, there has been only one other similar claim.  Valtonen,
Teerikorpi \& Argue \cite{Valtonen} found a tendency for compact satellites 
to be aligned with the major axes of highly--inclined disk galaxies;
however, their sample consisted of only 7 host galaxies. 
Although it is extremely tempting to accept its veracity based upon
an intuitive sense that planar alignment of satellites is more
dynamically
sensible than polar alignment, it is clear that the Brainerd 
\cite{TGB:holmberg}
result is badly in need of independent confirmation.

Sales \& Lambas \cite{SL} used a data set of very similar size (and in
one case identical selection criteria) to that of Brainerd \cite{TGB:holmberg}
yet did not detect any anisotropy in the satellite distribution when
they analysed their entire sample.  Why this is the case remains a
mystery at the moment, but it may be attributable to a combination
of two things.  First, the velocity errors in the 2dFGRS are typically
larger than those in the SDSS ($\sim 85$~\kms versus $\sim 20$~\kms
to $\sim 30$~\kms).  At some level, this would lead to a higher 
fraction of interlopers (i.e., false satellites) in the Sales \& Lambas
\cite{SL} sample than in the Brainerd \cite{TGB:holmberg} samples.  
Second, van den
Bosch et al.\ \cite{vandenBosch:2004a} found that when they combined mock redshift 
surveys with the 2dFGRS, there was a clear absence of satellites at
small projected radii in the 2dFGRS.  Since the majority of the
anisotropy seen by Brainerd \cite{TGB:holmberg} appears to come primarily from
small scales, it could be that Sales \& Lambas \cite{SL} simply had
too few pairs of hosts and satellites at small separations to detect
the anisotropy.
Any lack of host--satellite pairs in the 2dFGRS data, however, does
not explain why a Holmberg effect was detected by Sales \& Lambas
\cite{SL}
when they restricted their analysis to host--satellite pairs with
$|dv| < 160$~\kms.  When Brainerd \cite{TGB:holmberg} imposed the same restriction
on her sample 1 (i.e., the sample selected using the Sales \& Lambas
\cite{SL} selection criteria), 
she found that the satellites with $|dv| < 160$~\kms 
displayed an anisotropy that was identical to that of the full
sample: a clear alignment of the satellites with the host major axes.
The cause of this discrepancy is not at all obvious.  It may in part be
attributable to the fact that a value of $|dv| = 160$~\kms is
comparable to the error in a typical measurement of $|dv|$ for
hosts and satellites in the 2dFGRS.  Also, work by van den Bosch 
et al.\ \cite{vandenBosch:2004b} 
suggests that the interloper fraction is substantially
higher for host--satellite pairs with low values of $|dv|$ than it
is for host--satellite pairs with high values of $|dv|$.  It could,
therefore, be possible that the Sales \& Lambas \cite{SL} sample with
$|dv| < 160$~\kms is heavily contaminated with interlopers and some
strange, unknown selection bias is giving rise to their signal. 

Finally, it is worth notating that not only are the observational
conclusions about the distribution of satellite galaxies 
particularly muddy
at the moment, so
too are the theoretical conclusions.
Zaritsky et al.\ \cite{ZSFW} compared their observed Holmberg effect with
high--resolution CDM simulations and were unable to recover their
observations.  
Pe\~narrubia
et al.\ \cite{Penarrubia} investigated both polar and planar orbits of satellites
inside a massive,
flattened dark matter halo and found that the planar orbits 
decayed more quickly that the polar orbits.  They therefore 
suggest that such differences in orbital decay rates could be the origin of the
Holmberg effect. 
Abadi et al.\ \cite{Abadi}
suggest that the Holmberg effect could be caused by the cumulative
effects of accretion of satellites by the primary. 
However, Knebe et al.\ \cite{Knebe} found that the
orbits of satellites
of primary galaxies in cluster environments were located preferentially
within a cone of opening angle 40$^\circ$ (i.e., planar alignment,
not polar).  Since the structure of cold dark matter halos is
essentially independent of the mass scale of the halo
(e.g., \cite{Ben}, \cite{Tolya:1999}), the
implication of this result would be a preference for the satellites of
isolated galaxies to be aligned with the major axis of the host.
All of this in mind, perhaps the only
answer to the question ``Are either the Sales \& Lambas \cite{SL}
or Brainerd \cite{TGB:holmberg} observations of
anisotropic satellite distributions consistent with galaxy halos in 
a CDM universe?'' is, for now, ``Maybe''. 

\section{Summary}

There has been a long period of time over which it has been perfectly 
acceptable to write papers on investigations into the nature
of the dark matter halos of field galaxies that begin with a statement
along the lines of ``Although modern theories of galaxy formation
posit that all large galaxies reside within massive halos of dark
matter, the characteristic properties of those halos (e.g., mass,
radial extent, and shape) are not well--constrained by the current
observations''.   That time is now coming to an end.  The wealth
of data that has been acquired in recent years is truly beginning to
place strong, direct constraints on the dark matter
halos of field galaxies. 

Weak lensing and satellite dynamics have proven themselves to be
excellent probes of the gravitational potentials of large, bright
galaxies on physical scales $r \go 100~h^{-1}$~kpc.  
While one might be skeptical and discount the results
that come from one technique or the other, the fact that both are
yielding consistent constraints cannot be ignored.  Both weak 
lensing and satellite dynamics lead to statistical constraints on
the halo population as a whole, rather than constraints on any
one particular galaxy halo, and it is especially the acquisition
of extremely large data sets that has allowed these techniques to begin
to fulfill their promise of mapping out the gravitational potentials
associated with large, massive halos.  
Weak lensing and satellite dynamics have
inherent advantages and disadvantages, but since their systematic
errors and selection biases
are completely uncorrelated, they are extremely complementary
to each other.  At least at the moment, when strong constraints
are only just beginning to emerge from each technique, this 
complementarity is very reassuring.

Based upon my own critical, and hopefully unbiased, reading of
the recent literature, I think it is fair to say that, both
individually and in combination, weak lensing and satellite 
dynamics are pointing toward the following scenario for the
nature of large, bright field galaxies and their halos:

\begin{itemize}
\item The dark matter halos are well-characterized by NFW--type
objects in terms of their gravitational properties.  The dynamics
of satellite galaxies strongly prefer NFW halos to isothermal
halos.

\smallskip
\item The virial masses that are inferred for large field galaxies
are in good agreement with the predictions for galaxy--mass halos
in the context of cold dark matter.  Specifically, the virial
mass of the halo of
an ``average'' $\Lstar$ galaxy is in the range
$(8 - 10) \times 10^{11} h^{-1} \Msun$ when NFW profiles are fit
to the data.

\smallskip
\item There are clear differences in the depths of the
potential wells of the halos that surround galaxies of differing
morphology and differing intrinsic luminosity.  Specifically,
the virial
masses of the halos of
$\Lstar$ ellipticals exceed those of $\Lstar$ spirals by a factor
of at least 2.  The actual value of the mass
excess depends upon details of the data and its analysis.
In addition, the virial
masses of the halos of high luminosity galaxies exceed those of
low luminosity galaxies.  Again, however, the amount by which they
differ depends upon details of the data and its analysis.

\smallskip
\item Averaged over all galaxies with 
$L \go \Lstar$, the mass--to--light ratio computed on scales
larger than the optical radii of the galaxies
is, at most, weakly--dependent
upon the luminosity of the galaxy.  At the 2$\sigma$ level, the
mass--to--light ratio of the average galaxy with $L \go \Lstar$ 
is consistent with a constant value.

\smallskip
\item The dark matter halos are flattened, rather than spherical,
and the degree of flattening on large scales ($\sim 100$~kpc to
$\sim 200$~kpc) is consistent with the predictions of cold dark
matter.
\end{itemize}

It is worth noting that the above list comes from  quite a diverse
set of data.  In particular, the data are spread over a wide
range in redshift.  With the exception of preliminary data from DEEP2,
the satellite dynamics studies have median redshifts of $z_{\rm med} 
\sim 0.07$.  The weak lenses in the SDSS data have a median redshift
of $z_{\rm med} \sim 0.16$  and the weak lenses in the RCS and
COMBO--17 data have considerably higher redshifts,
$z_{\rm med} \sim 0.4$.  Since it is clear that the field galaxy
population has evolved since $z \sim 0.5$, it is not entirely fair
to lump the results from all of these studies together, and I think
the big challenge to the weak lensing community in particular will
be to eventually place constraints on the evolution of field galaxies
and their halos from, say, $z \sim 1$ to the present.

Nevertheless, I think we have reached a particularly gratifying time
in which we are really being able to measure some of the fundamental
properties of dark mater halos on physical scales that extend well
beyond the visible images of the galaxies at their centers. A
remarkably consistent picture of the large--scale gravitational 
properties of the halos is emerging from the observations and, at
least for now, that picture seems entirely in accord with a
cold dark matter universe. 


\begin{theacknowledgments}
I am deeply indebted to Henk Hoekstra, Martina Kleinheinrich,
and Erin Sheldon for their help
with the preparation of numerous
figures at a time when they all had much more
important things do to, and to Tom Peterson, without whose indulgence
and encouragement this article would probably never have been written.
Support under NSF contracts AST--0098572 and AST--0406844 is also gratefully
acknowledged. 
\end{theacknowledgments}





\end{document}